\documentclass[aps,prd,floatfix,showpacs]{revtex4}
\usepackage{graphics}
\usepackage{graphicx}
\usepackage{amsmath,amssymb}
\usepackage{tabularx}
\usepackage[english,activeacute] {babel}

\begin{document}

\title{FCNC, CP violation and implications for some rare decays in an $SU(4)_L\otimes U(1)_X$ extension of the standard model} 

\author{Alejandro Jaramillo and Luis A. S\'anchez}

\affiliation{Escuela de F\'\i sica, Universidad Nacional de Colombia,
A.A. 3840, Medell\'\i n, Colombia}

%%%%%%%%%%%%%%%%%%%%%%%%%%%%%%%%%%%%%%%%%%%%%%%%%%%%%%%%%%%%
\begin{abstract}
Extensions of the standard model (SM) with gauge symmetry $SU(3)_c\otimes SU(4)_L\otimes U(1)_X$ (3-4-1 extensions) where anomaly cancellation takes place between the fermion families (three-family models) predict the existence of two new heavy neutral gauge bosons which transmit flavor changing neutral currents (FCNC) at tree-level. In this work, in the context of a three-family 3-4-1 extension which does not contain particles with exotic electric charges, we study the constraints coming from neutral meson mixing on the parameters of the extension associated to tree level FCNC effects. Taking into account experimental measurements of observables related to $K$ and $B$ meson mixing and including new CP-violating phases, we study the resulting bounds for angles and phases in the mixing matrix for the down-quark sector, as well as the implications of these bounds for the modifications in the amplitudes of the clean rare decays $K^+ \to \pi^+ \bar \nu \nu$, $K_{ L} \to \pi^0\nu\bar\nu$, $K_L \to \pi^0 l^+ l^-$ ($l=e, \mu$) and $B_{d/s}\to \mu^+ \mu^- $.
\end{abstract}
%%%%%%%%%%%%%%%%%%%%%%%%%%%%%%%%%%%%%%%%%%%%%%%%%%%%%%%%%%%%


\maketitle
%%%%%%%%%%%%%%%%%%%%%%%%%%%%%%%%%%%%%%%%%%%%%%%%%%%%%%%%%%%%
\section{\label{sec:intr}Introduction}
%%%%%%%%%%%%%%%%%%%%%%%%%%%%%%%%%%%%%%%%%%%%%%%%%%%%%%%%%%%%
Flavor violating couplings of ordinary fermions to extra neutral gauge bosons and to new scalar fields arise in many extensions of the standard model (SM). Two simple and interesting examples are the 3-3-1 model in which the SM gauge symmetry is enlarged to $SU(3)_c\otimes SU(3)_L\otimes U(1)_X$ \cite{331-1,331-2,Liu:1993gy,Promberger:2007py,   
Cabarcas:2007my}, and the 3-4-1 extension where the enlargement is done to the gauge group $SU(3)_c\otimes SU(4)_L\otimes U(1)_X$ \cite{341-1,341-2,Ponce:2006vw,Nisperuza:2009xm,Villada:2009iu}. These extensions share the important feature of addressing the problem of the number of fermion families in nature in the sense that anomaly cancellation among the families (three-family models) takes place only if we have an equal number of left-handed triplets and antitriplets (in the 3-3-1 model) or an equal number of $4$-plets and $4^*$-plets (in the 3-4-1 extension), taking into account the color degree of freedom. As a consequence, the number of fermion families $N_f$ must be divisible by the number of colors $N_c$ of $SU(3)_c$. Moreover, since $SU(3)_c$ asymptotic freedom requires $N_c < 5$, it follows that $N_f = N_c = 3$. Cancellation of chiral anomalies among the families thus forces one family of quarks, in the weak basis, to transform differently from the other two under $SU(4)_L\otimes U(1)_X$. As a result, the chiral couplings of quarks to the new neutral gauge bosons are, in general, nonuniversal and, when rotating to the quark mass eigenstate basis, flavor changing neutral currents (FCNC) induced by fermion mixing arise.

In particular, the 3-4-1 extension predicts the existence of two heavy neutral gauge bosons $Z^\prime$ and $Z^{\prime \prime}$ which, in general, mix up with the ordinary $Z$ boson of the SM. In contrast with the SM where FCNC processes occur only at the loop level, in the 3-4-1 extension these new gauge bosons can transmit FCNC at tree level and, consequently, the study of these processes can set stringent bounds on the $Z^\prime$ and $Z^{\prime \prime}$ masses and mixing. Moreover, since in general each flavor couples to more than one Higgs 4-plet, FCNC coming from the scalar sector can also be present.

For simplicity, in this work we will restrict ourselves to 3-4-1 extensions without exotic electric charges in the fermion sector, that is, without electric charges different from $\pm 2/3$ and $\pm 1/3$ for exotic quarks and different from $0$ and $\pm 1$ for exotic leptons. The systematic analysis of the 3-4-1 gauge theory carried out in \cite{Ponce:2006vw} has shown that the restriction to fermion field representations with only ordinary electric charges allows for eight different anomaly free extensions. Four of them are three-family models and can be classified according to the values of the coefficients $b$ and $c$ which appear in the most general expression for the electric charge generator in $SU(4)_L\otimes U(1)_X$ [see Eq.~(\ref{eq:Q341}) below]. The allowed simultaneous values for these coefficients are $b= c = 1$ and $b = 1,\; c = -2$. Two of the four three-family models belong to the $b = c = 1$ class; the other two belong to the $b = 1,\; c = -2$ class.

In 3-4-1 extensions without exotic electric charges the $Z-Z^\prime-Z^{\prime \prime}$ mixing can be constrained to occur between $Z$ and $Z^\prime$ only, which leaves $Z^{\prime \prime} \equiv Z_3$ as a heavy mass eigenstate \cite{341-2, Ponce:2006vw, Nisperuza:2009xm, Villada:2009iu}. The diagonalization of the $Z-Z^\prime$ mass matrix produces a light mass eigenstate $Z_1$ which can be identified as the neutral gauge boson of the SM, and a heavy $Z_2$. After the breakdown of the 3-4-1 symmetry down to $SU(3)_c\otimes U(1)_{Q}$, and since we have one family of quarks transforming differently from the other two under the gauge group, one important difference between the aforementioned two classes of 3-4-1 extensions appears: even thought in both classes the $Z_1$ current remains flavor diagonal, in the $b=c=1$ class the new $Z_2$ gauge boson couples nonuniversally to ordinary quarks thus transmitting tree-level FCNC at low energies, while the couplings to $Z_3$ are universal and $Z_3$ couples only to exotic fermions. In the $b=1, \; c=-2$ class, instead, it is the new $Z_3$ gauge boson the responsible for this effect because couples nonuniversally to ordinary quarks, while the $Z_2$ current remains flavor diagonal and universal.

The study of the bounds on the $Z_2$ and $Z_3$ masses coming from electroweak precision data and from FCNC processes has shown that extensions for which $b=1, \; c=-2$ are preferred in the sense that they give lower bounds on these masses smaller than the bounds predicted by extensions in the $b=c=1$ class \cite{Nisperuza:2009xm}. In fact, in the latter class a $\chi^2$ fit to $Z$-pole observables and atomic parity violation data produces $m_{Z_2}\gtrsim 2$ TeV, a bound that is further increased to $m_{Z_2}\gtrsim 11$ TeV when the constraints coming from neutral meson mixing in the study of the tree level FCNC effects are taken into account. In the $b=1, \; c=-2$ class, instead, the same fit gives the lower bound $m_{Z_2}\gtrsim 0.8$ TeV and the analysis of the constraints arising from neutral meson mixing provides $m_{Z_3}\gtrsim 6.5$ TeV. This means that extensions in the $b=1, \; c=-2$ class have a better chance to be tested at the LHC facility or further at the ILC. Here however we must point out that, as it has been noted repeatedly in the literature, the bounds from neutral meson mixing are always clouded by the lack of knowledge of the entries in the quark mixing matrices $V_L^u$ and $V_L^d$ involved in the new physics contribution, which forces to adopt well motivated but ad hoc textures for these matrices.

In this work, in the context of a 3-4-1 extension belonging to the $b=1, \; c=-2$ class (the so-called ``Model F'' in Ref.~\cite{Ponce:2006vw}), we re-examine the issue of tree level FCNC transmitted by the new $Z_3$ gauge bosons but, due to the uncertainties associated with our ignorance of the quark mixing matrices, we will not search for bounds on the $Z_3$ mass, but rather we set this mass at fixed values in the range $1-5$ TeV (to be justified below) and search for information about the size of angles and phases in the $V_L^d$ mixing matrix. To this purpose we will use several well measured $\Delta F=2$ ($F=S,B$) observables in the down sector, namely $\Delta M_K$, $\Delta M_{d/s}$, $\varepsilon_K$, and $\sin \Phi_d$, associated to the $K_L-K_S$ and the $B_{d/s}^{0}-\bar B_{d/s}^{0}$ mass differences, the Kaon CP-violation parameter and the $B_{d}^{0}-\bar B_{d}^{0}$ mixing phase, respectively. We will also study the implications of the obtained bounds for the modifications in the amplitudes of the clean rare decays $K^+ \to \pi^+  \nu \bar \nu$, $K_L \to \pi^0 \nu \bar \nu$, $B_{d/s} \to \mu^+ \mu^-$ and $K_L \to \pi^0 l^+ l^-$ ($l=e,\mu$). In this context our mail goal will be to obtain upper and lower bounds for the corresponding branching ratios (BR) and to study their compatibility both with the experimental data and the SM predictions. 
 
Another well known FCNC rare process that deserves attention due to its sensitivity to new physics, is the radiative decay $b \to s\gamma$. In general, this decay receives contributions both from the new charged and neutral gauge bosons and from the new scalar fields and can be used to put limits on these sectors of the 3-4-1 extension. As can be seen from the 3-4-1 scalar structure in Eq.~(\ref{vev}) below, and as it is done in the 3-3-1 model, the scalar contributions can be accounted for by an effective two-Higgs-doublet model. This, however, demands the identification of the physical scalar fields and their couplings and, therefore, the diagonalization of the full scalar sector. It is then clear that the study of this decay requires a particular and dedicated analysis. Moreover, as in the SM, $b \to s\gamma$ is also a loop process in the 3-4-1 construction and since, as declared above, we are mainly interested in the analysis of the more stringent tree level FCNC processes that receive contributions from $Z_3$ exchange, the detailed study of this decay in the context of the present SM extension will be postponed to a future work.
 
This paper is organized as follows. In Sec. \ref{sec:sec2}, we introduce the 3-4-1 extension to be considered and present its most important features for our purposes. In Secs.~\ref{sec:sec3} and~\ref{sec:sec4}, we study the theoretical expression for the observables associated to neutral meson mixing and for the rare decays that will be analyzed, in such a way that the tree level FCNC contributions of the extra neutral gauge boson enter as corrections to the SM expressions. In Sec.~\ref{sec:sec5}, we numerically evaluate the theoretical expressions obtained in the two previous sections, from which some information on structure of the $V^d_L$ mixing matrix can be obtained and the implications for the rare decays can be calculated. In the last Section we summarize our results.
%%%%%%%%%%%%%%%%%%%%%%%%%%%%%%%%%%%%%%%%%%%%%%%%%%%%%%%%%%%%

%%%%%%%%%%%%%%%%%%%%%%%%%%%%%%%%%%%%%%%%%%%%%%%%%%%%%%%%%%%%
\section{\label{sec:sec2}The 3-4-1 extension} 
Let us begin by introducing the most relevant features of the 3-4-1 extension we are interested in. Several details of their phenomenology have been already worked out in \cite{Villada:2009iu}.

The extension is based on the local $SU(3)_c\otimes SU(4)_L\otimes U(1)_X$ gauge symmetry which contains $SU(3)_c\otimes SU(2)_L\otimes U(1)_Y$ as a subgroup, and belongs to the $b=1,\; c= -2$ class, where $b$ and $c$ are parameters appearing in the most general expression for the electric charge generator in $SU(4)_L\otimes U(1)_X$

\begin{equation}\label{eq:Q341}
Q=aT_{3L}+\frac{1}{\sqrt{3}}bT_{8L}+ \frac{1}{\sqrt{6}}cT_{15L}+ XI_4, 
\end{equation} 
where $T_{iL}=\lambda_{iL}/2$, with $\lambda_{iL}$ the Gell-Mann matrices
for $SU(4)_L$ normalized as  Tr$(\lambda_i\lambda_j)=2\delta_{ij}$,
$I_4={\rm diag}(1,1,1,1)$ is the diagonal $4\times 4$ unit matrix, and $a=1$ gives the usual isospin of the electroweak interaction.

The anomaly-free fermion content of this extension has been discussed in Ref.~\cite{Ponce:2006vw} and is given in Table \ref{tab:1} where $i=1,2$ and $\alpha=1,2,3$ are generation indexes. The numbers inside brackets match the $SU(3)_c, SU(4)_L$ and $U(1)_X$ quantum numbers, respectively. $U_i$ and $U_3$ are exotic quarks of electric charge $2/3$, $D_i$ and $D_3$ are also exotic quarks with electric charge $-1/3$, while $E^-_\alpha$ and $N^0_\alpha$ are exotic leptons with electric charges $-1$ and $0$, respectively.

\begin{table}
\caption{\label{tab:1}Anomaly-free fermion content.}
\begin{center}
\begin{ruledtabular}
\begin{tabular}{ccccc}
$Q_{iL}=\left(\begin{array}{c}d_i\\u_i\\U_i\\D_i \end{array}\right)_L$ &
$d^c_{iL}$ & $u^c_{iL}$ & $U^c_{iL}$ & $D^{c}_{iL}$ \\
$[3,4^*,\frac{1}{6}]$ & $[3^*,1,\frac{1}{3}]$ & $[3^*,1,-\frac{2}{3}]$
& $[3^*,1,-\frac{2}{3}]$ & $[3^*,1,\frac{1}{3}]$ \\ \hline
$Q_{3L}=\left(\begin{array}{c}u_3\\d_3\\D_3\\U_3
\end{array}\right)_L$ &
$u^c_{3L}$ & $d^c_{3L}$ & $D^c_{3L}$ & $U^{c}_{3L}$ \\
$[3,4,\frac{1}{6}]$ & $[3^*,1,-\frac{2}{3}]$ & $[3^*,1,\frac{1}{3}]$ & $[3^*,1,\frac{1}{3}]$ & $[3^*,1,-\frac{2}{3}]$ \\ \hline
$L_{\alpha L}=\left(\begin{array}{c}\nu _{e\alpha }^{0} \\e_{\alpha }^{-} \\E_{\alpha }^{-} \\N_{\alpha }^{0} \end{array}\right)_L $ &
$e^+_{\alpha L}$ & $E^+_{\alpha L}$ & $$ & $$ \\
$[1,4,-\frac{1}{2}]$ & $[1,1,1]$ & $[1,1,1]$ & $$ & $$\\
\end{tabular}
\end{ruledtabular}
\end{center}
\end{table}

The symmetry breaking occurs in three steps
\begin{eqnarray}\nonumber
SU(3)_c\otimes SU(4)_L\otimes & U(1)_X & \\ \nonumber
& \stackrel{V^\prime}{\longrightarrow}
& SU(3)_c\otimes SU(3)_L\otimes U(1)_Z \\ \nonumber
& \stackrel{V}{\longrightarrow} & SU(3)_c\otimes SU(2)_L\otimes U(1)_Y \\ \label{break}
& \stackrel{v+v^\prime}{\longrightarrow} & SU(3)_c\otimes U(1)_Q,
\end{eqnarray}
a task that is done by the following four Higgs fields with vacuum expectation values (VEV) aligned as

\begin{eqnarray}\nonumber
\left\langle \phi _{1}^{T}\right\rangle & =\left\langle \left( \phi
_{1}^{0},\phi _{1}^{+},\phi _{1}^{\prime +},\phi _{1}^{\prime 0}\right)
\right\rangle =\left( v,0,0,0\right) \sim \left[ 1,4^{\ast },1/2\right] , \\ \nonumber
\left\langle \phi _{2}^{T}\right\rangle & =\left\langle \left( \phi
_{2}^{-},\phi _{2}^{0},\phi _{2}^{\prime 0},\phi _{2}^{\prime -}\right)
\right\rangle =\left( 0,v^{\prime},0,0\right) \sim \left[ 1,4^{\ast },-1/2
\right] , \\ \nonumber
\left\langle \phi _{3}^{T}\right\rangle & =\left\langle \left( \phi
_{3}^{-},\phi _{3}^{0},\phi _{3}^{\prime 0},\phi _{3}^{\prime -}\right)
\right\rangle =\left( 0,0,V,0\right) \sim \left[ 1,4^{\ast },-1/2\right], \\ \label{vev}
\left\langle \phi _{4}^{T}\right\rangle & =\left\langle \left( \phi
_{4}^{0},\phi _{4}^{+},\phi _{4}^{\prime +},\phi _{4}^{\prime 0}\right)
\right\rangle =\left( 0,0,0,V^{\prime}\right) \sim \left[ 1,4^{\ast },1/2
\right].
\end{eqnarray}

This scalar structure consistently gives masses for all the gauge bosons and it is also enough to produce the observed fermion mass spectrum for the charged fermion sector (quarks and leptons) provided the hierarchy $V,\, V^\prime >> v,\; v^\prime \simeq 174$~GeV is satisfied, where $V^\prime$ and $V$ set the mass scales for exotic fields \cite{Villada:2009iu}.

The gauge couplings $g_4$ and $g_X$, associated with the groups $SU(4)_L$ and $U(1)_X$, respectively, are defined through the covariant derivative for 4-plets as: $iD^\mu = i\partial^\mu - g_4 \lambda_{L\alpha}A^\mu_\alpha/2 - g_X X B^\mu$.
When the 3-4-1 symmetry is broken to the SM, we get the gauge matching conditions
\begin{eqnarray}
g_4 = g, \qquad&\mbox{and}&\qquad\frac{1}{g^{\prime 2}}=\frac{1}{g^2_X}+\frac{1}{g^2}, \label{matchA}
\end{eqnarray}
where $g$ and $g^\prime$ are the gauge coupling constants of the $SU(2)_L$ and $U(1)_Y$ groups of the SM, respectively.

Clearly, FCNC in this extension can arise from the mixing of ordinary and exotic fermions. These FCNC and violation of the unitarity of the CKM mixing matrix can be avoided by the introduction of the following $Z_2$ discrete symmetry which assigns charges $q_{\mathrm{Z}}$ to the fields as \cite{Villada:2009iu}
\begin{eqnarray}\nonumber
q_{\mathrm{Z}}(Q_{\alpha L},u_{\alpha L}^{c},d_{\alpha L}^{c},L_{\alpha
L},e_{\alpha L}^{+},\phi _{1},\phi _{2})&=& 0, \\ \label{Zcharge}
q_{\mathrm{Z}}(U_{\alpha L},D_{\alpha L}^{c},E_{\alpha L}^{+},\phi _{3},\phi
_{4})&=& 1,
\end{eqnarray}
where $\alpha=1,2,3$ is a family index as above.

The gauge invariance and the $Z_2$ symmetry allow for the following Yukawa Lagrangian in the quark sector
\begin{eqnarray}\nonumber
\bigskip \mathcal{L}_{Y}^{Q} &=&\sum_{i=1}^2Q_{iL}^{T}C[\phi
_{2}^{\ast }\sum_{\alpha=1}^3h_{i\alpha }^{u}u_{\alpha
L}^{c}+\phi _{1}^{\ast }\sum_{\alpha=1}^3h_{i\alpha
}^{d}d_{\alpha L}^{c}+\phi _{3}^{\ast }\sum_{\alpha=1}^3h_{i\alpha }^{U}U_{\alpha L}^{c} \\ \nonumber
&&+\phi _{4}^{\ast }\sum_{\alpha=1}^3h_{i\alpha }^{D}D_{\alpha
L}^{c}]+Q_{3L}^{T}C[\phi _{1}\sum_{\alpha=1}^3h_{i\alpha
}^{u}u_{\alpha L}^{c}+\phi _{2}\sum_{\alpha=1}^3h_{3\alpha
}^{d}d_{\alpha L}^{c} \\ \label{lagquark}
&&+\phi _{4}\sum_{\alpha=1}^3h_{3\alpha }^{U}U_{\alpha
L}^{c}+\phi _{3}\sum_{\alpha=1}^3h_{3\alpha }^{D}D_{\alpha
L}^{c}]+h.c.,
\end{eqnarray}
where the $h^\prime s$ are Yukawa couplings and $C$ is the charge conjugation operator. From this Lagragian we get, for the up- and down-type quarks in the basis $(u_{1},u_{2},u_{3},U_{1},U_{2},U_{3})$ and $(d_{1},d_{2},d_{3},D_{1},D_{2},D_{3})$, respectively, $6\times 6$ block diagonal mass matrices of the form

\begin{equation}\label{massquark}
M_{uU}=\left(\begin{array}{cc}M_{3\times 3}^{u} & 0 \\
0 & M_{3\times 3}^{U}
\end{array}
\right) \quad \mathrm{and} \quad M_{dD}=\left(\begin{array}{cc}M_{3\times 3}^{d} & 0 \\
0 & M_{3\times 3}^{D}
\end{array}
\right),
\end{equation}
where, for $V\sim V^\prime$ and $v\sim v^\prime$, the entries in the submatrices are
\begin{eqnarray}\nonumber
M^u_{\alpha \beta }\simeq h_{\alpha \beta }^{u}v &\quad{\mbox{and}}&\quad M^U_{\alpha \beta
}\simeq h_{\alpha \beta }^{U}V; \\ \label{submq}
M^d_{\alpha \beta }\simeq h_{\alpha \beta }^{d}v &\quad{\mbox{and}}&\quad M^D_{\alpha \beta
}\simeq h_{\alpha \beta }^{D}V.
\end{eqnarray}

For the charged leptons we get the Yukawa terms
\begin{equation}\label{laglepton}
\bigskip \mathcal{L}_{Y}^{L}=\sum_{\alpha=1}^3\sum_{\beta=1}^3L_{\beta L}^{T}C[\phi _{2}h_{\alpha \beta }^{e}e_{\beta L}^{+}+\phi
_{3}h_{\alpha \beta }^{E}E_{\beta L}^{+}]+h.c.
\end{equation}
From this equation we find a block diagonal mass matrix in the basis $(e_{1},e_{2},e_{3},E_{1},E_{2},E_{3})$ given by
\begin{equation}\label{masslepton}
M_{eE}=\left(\begin{array}{cc}
M_{3\times 3}^{e} & 0 \\
0 & M_{3\times 3}^{E}
\end{array}
\right),
\end{equation}
where now the entries in the submatrices are
\begin{equation}\label{subml}
M^e_{\alpha \beta }=h_{\alpha \beta }^{e}v^{\prime} \quad{\mbox{and}}\quad M^E_{\alpha \beta
}=h_{\alpha \beta }^{E}V.
\end{equation}
The mass matrices in (\ref{massquark}) and (\ref{masslepton}) exhibit a simple mass splitting between ordinary and exotic charged fermions, and show that all the charged fermions in this extension acquire masses at the three level. Clearly, by a judicious tuning of the Yukawa couplings a consistent mass spectrum in the ordinary charged sector can be obtained. In the exotic charged sector all fermions acquire masses at the scale $V\sim V^\prime \gg v_{EW}=174$ GeV. We also remark that the tensor product form of the mass matrices $M_{uU}$ and $M_{dD}$ in (\ref{massquark}) implies that they are diagonalized by unitary matrices which are themselves tensor products of unitary matrices. So, the discrete $Z_2$ symmetry also avoids violation of unitarity of the CKM mixing matrix. The neutral leptons $\nu_{e\alpha}^{0}$ and $N_{e\alpha}^{0}$ ($\alpha=1,2,3$) remain massless as far as we use only the original fermion fields shown in Table \ref{tab:1}. However, their masses and mixing can be implemented by introducing one or more Weyl singlet states $N_{L,n}^{0}\sim [1,1,0],\; n =1,2,...$, without violating our assumptions, neither the anomaly constraint relations, because singlets with no $X$-charges are as good as not being present as far as anomaly cancellation is concerned.

A look at the Yukawa Lagrangians in Eqs.~(\ref{lagquark}) and (\ref{laglepton}) shows that each flavor couples to more than one Higgs 4-plet and, consequently, scalar mediated FCNC also arise. However, since all the processes we will consider in this work involve external light quarks and leptons, these contributions are suppressed by the small Yukawa couplings associated to these fields. Notice that, for these processes, this will also be the case even in the absence of the discrete $Z_2$ symmetry introduced in Eq.~(\ref{Zcharge}) because the Yukawa couplings will now be suppressed by small mixing angles. In the charged boson sector there are SM-like $W^\pm$ gauge bosons that do not mix with the other charged bosons and acquire a squared mass $M_{W^{\pm}}^{2}=(g_{4}^{2}/2)(v^{2}+v^{\prime 2})$ so that, with $M_{W}= 80.399 \pm 0.023$ GeV \cite{Nakamura:2010zzi}, we have $\sqrt{v^{2}+v^{\prime 2}} \simeq v_{\rm{EW}} = 174$ GeV. The ten remaining physical charged gauge bosons, namely: $K^{\pm}$, $V^{\pm}$, $Y^{\pm}$, $X^{0}(X^{\prime 0})$, and $K^{0}(K^{\prime 0})$, acquire masses at the large scale $V^\prime \sim V$ and, at tree level, couple always to at least one exotic fermion \cite{Villada:2009iu}. This means that, for low energy processes involving ordinary fermions, the contribution of the new charged gauge bosons will be present only at loop level. Thus, we expect that the tree level FCNC effects transmitted by the exotic neutral gauge bosons will dominate.

Our main interest, therefore, is in the neutral gauge boson sector which consists of four physical fields: the massless photon $A_\mu$ and three massive gauge bosons which come from the diagonalization of the mixing $Z_\mu-Z^\prime_\mu-Z^{\prime \prime}_\mu$. In terms of the electroweak basis, they are given by

\begin{eqnarray} \nonumber
A^\mu&=&S_W A_3^\mu \nonumber \\
& & + C_W\left[\frac{T_W}{\sqrt{3}}\left(A_8^\mu-
2\frac{A_{15}^\mu}{\sqrt{2}}\right)+(1-T_W^2)^{1/2}B^\mu\right]\; , \nonumber \\  
Z^\mu&=& C_W A_3^\mu \nonumber \\
& & - S_W\left[\frac{T_W}{\sqrt{3}}\left(A_8^\mu-
2\frac{A_{15}^\mu}{\sqrt{2}}\right)+(1-T_W^2)^{1/2}B^\mu\right] \; , \nonumber \\ \nonumber
Z'^\mu&=&\frac{1}{\sqrt{3}}(1-T_W^2)^{1/2}\left(A_8^\mu-
2\frac{A_{15}^\mu}{\sqrt{2}}\right)-T_W B^\mu, \\ \label{fzzpb}
Z''^\mu&=& 2A_8^\mu /\sqrt{6}+A_{15}^\mu/\sqrt{3}.
\end{eqnarray}
\noindent

Since we are interested in the low energy phenomenology, we can choose $V \simeq V^\prime$. If we also take $v\simeq v^\prime$, the current $Z^{\prime \prime \mu}\equiv Z^\mu_3$ decouples from the other two and acquires a squared mass $M^2_{Z_3}=(g^2_4/2)(V^2+v^2)$ \cite{341-2, Ponce:2006vw, Nisperuza:2009xm, Villada:2009iu}. This fact produces an enormous simplification in the study of the low energy deviations of the $Z$ couplings to the SM families which now come from the diagonalization of the mass matrix  
\begin{equation}\label{remaining}
M_{(Z,Z^\prime)} = \frac{g_{4}^{2}}{C_{W}^{2}}\left(
\begin{array}{cc}
v^{2} & \delta v^{2}S_{W} \\
\delta v^{2}S_{W} & \qquad \frac{\delta ^{2}}{S_{W}^{2}}\left(
V^{2}C_{W}^{4}+v^{2}S_{W}^{4}\right)
\end{array}
\right),
\end{equation}
where $\delta =g_{X}/g_{4}$, and $S_{W}=\delta /\sqrt{2\delta ^{2}+1}$ and $C_{W}$ are the sine and cosine of the electroweak mixing angle, respectively. The corresponding mass eigenstates are: 
$Z_{1}^{\mu } = Z^{\mu }\cos \theta +Z^{\prime \mu}\sin \theta$ and 
$Z_{2}^{\mu } =-Z^{\mu }\sin \theta +Z^{\prime \mu}\cos \theta$,
where the mixing angle $\theta$ is given by
\begin{equation} \label{tanb} 
\tan(2\theta) =  \frac{v^2 S_W^2 \sqrt{C_{2W}}}
{v^2(1+S_W^2)^2 + V^2 C_W^4 - 2v^2},
\end{equation}
\noindent 
with $C_{2W}=C_{W}^{2}-S_{W}^{2}$. Since $V>>v$ this mixing angle is expected to be very small.

%%%%%%%%%%%%%%%%%%%%%%%%%%%%%%%%%%%%%%%%%%%%%%%%%%%%%%%

\subsection{Neutral Currents}
The Lagrangian for the neutral currents $J_{\mu}(\mathrm{EM})$, $J_{\mu}(Z)$, 
$J_{\mu}(Z^{\prime})$, and $J_{\mu}(Z^{\prime \prime})$, in the basis in which all the fields are gauge eigenstates, is given by

\begin{eqnarray}\nonumber
-\mathcal{L}^{\mathrm{NC}}&=&eA^{\mu }J_{\mu }(\mathrm{EM})+\frac{g_{4}}{C_{W}}Z^{\mu}J_{\mu}(Z)+g_{X}Z^{\prime\mu}J_{\mu}(Z^{\prime}) \\ \label{LNC}
&&+\frac{g_{4}}{2\sqrt{2}}Z^{\prime \prime}J_{\mu}(Z^{\prime \prime}),
\end{eqnarray}
where $e=gS_{W}=g_{X}C_{W}\sqrt{1-T_{W}^{2}}>0$. Calling $q_f$ the electric charge of the fermion $f$ in units of $e$, the currents are given by

\begin{eqnarray}\nonumber
J_{\mu}(\mathrm{EM}) &=&\frac{2}{3}[\bar{u}_{3}\gamma _{\mu}u_{3}+\bar{U}
_{3}\gamma _{\mu}U_{3}+\sum_{i=1}^2(\bar{u}_{i}\gamma _{\mu}u_{i}+\bar{U}
_{i}\gamma _{\mu}U_{i})] \\ \nonumber
&&-\frac{1}{3}[\bar{d}_{3}\gamma _{\mu }d_{3}+\bar{D}_{3}\gamma
_{\mu }D_{3}+\sum_{i=1}^2(\bar{d}_{i}\gamma _{\mu }d_{i}+\bar{D}_{i}\gamma
_{\mu }D_{i})] \\ \nonumber
&&-\sum_{\alpha=1}^3(\bar{e}_{\alpha}^{-}\gamma _{\mu }e_{\alpha }^{-}+\bar{E}
_{\alpha }^{-}\gamma _{\mu }E_{\alpha}^{-}) \\ \label{emcurr}
&=& \sum_{f} q_{f}\bar{f}\gamma _{\mu }f,
\end{eqnarray}

\begin{eqnarray}\nonumber
J_{\mu}(Z)&=&J_{\mu ,L}(Z)-S_{W}^{2}J_{\mu }(\mathrm{EM}) \\ \nonumber
&=& \frac{1}{2}[\bar{u}_{3L}\gamma _{\mu }u_{3L}-\bar{
d}_{3L}\gamma _{\mu }d_{3L}-\sum_{i=1}^2(\bar{d}_{iL}\gamma _{\mu }d_{iL}-\bar{u}_{iL}\gamma _{\mu}u_{iL}) \\ \label{zeta}
&&+\sum_{\alpha=1}^3(\bar{\nu }_{\alpha L}\gamma _{\mu }\nu _{\alpha L}-\bar{e}
_{\alpha L}^{-}\gamma _{\mu }e_{\alpha L}^{-})]-S_{W}^{2}J_{\mu }(\mathrm{EM}),
\end{eqnarray}

\begin{eqnarray}\nonumber
J_{\mu}(Z^{\prime})&=&J_{\mu,L}(Z^{\prime})-T_{W}J_{\mu}(\mathrm{EM}) \\ \nonumber
&=& (2T_{W})^{-1}[T_{W}^{2}\bar{u}_{3L}\gamma
_{\mu }u_{3L}-T_{W}^{2}\bar{d}_{3L}\gamma _{\mu }d_{3L}
-\bar{D}_{3L}\gamma _{\mu }D_{3L} \\ \nonumber
&&+\bar{U}_{3L}\gamma _{\mu}U_{3L}-\sum_{i=1}^2(T_{W}^{2}\bar{d}_{iL}\gamma _{\mu }d_{iL}-T_{W}^{2}\bar{u}
_{iL}\gamma _{\mu }u_{iL} \\ \nonumber
&&-\bar{U}_{iL}\gamma _{\mu }U_{iL}+\bar{D}_{iL}\gamma _{\mu
}D_{iL})+\sum_{\alpha=1}^3(T_{W}^{2}\bar{\nu }_{\alpha L}\gamma _{\mu }\nu _{\alpha
L} \\ \label{zetap}
&&-T_{W}^{2}\bar{e}_{\alpha L}^{-}\gamma _{\mu }e_{\alpha L}^{-}-\bar{E}_{\alpha L}^{-}\gamma _{\mu }E_{\alpha L}^{-}+\bar{N}
_{\alpha L}^{0}\gamma _{\mu }N_{\alpha L}^{0})]-T_{W}J_{\mu}(\mathrm{EM}),
\end{eqnarray}
and
\begin{eqnarray}\nonumber
J_{\mu}(Z^{\prime \prime}) &=&\sum_{i=1}^2(-\bar{d}_{iL}\gamma _{\mu}d_{iL}-
\bar{u}_{iL}\gamma _{\mu }u_{iL}+\bar{U}_{iL}\gamma _{\mu }U_{iL}+\bar{D}_{iL}\gamma _{\mu }D_{iL}) \\ \nonumber
&&+\bar{u}_{3L}\gamma _{\mu
}u_{3L}+\bar{d}_{3L}\gamma _{\mu }d_{3L}-\bar{D}_{3L}\gamma _{\mu }D_{3L}-\bar{U}_{3L}\gamma _{\mu}U_{3L} \\ \label{zprimaprima}
&&+\sum_{\alpha=1}^3(\bar{\nu }_{\alpha L}\gamma _{\mu }\nu _{\alpha L}+\bar{e}
_{\alpha L}^{-}\gamma _{\mu }e_{\alpha L}^{-}-\bar{E}_{\alpha L}^{-}\gamma _{\mu }E_{\alpha L}^{-}-\bar{N}
_{\alpha L}^{0}\gamma _{\mu }N_{\alpha L}^{0}), 
\end{eqnarray}
Eq.~(\ref{zeta}) shows that we can identify $Z_\mu$ as the neutral gauge boson of the SM because $J_{\mu}(Z)$ is just the generalization of the SM neutral current. Moreover, from (\ref{zetap}) it is straightforward to see that the neutral gauge boson $Z_\mu^\prime$ does not transmit FCNC at low energy since it couples diagonally and universally to ordinary fermions.

Notice, on the other hand, that $J_{\mu}(Z^{\prime \prime})$ is a pure left-handed current and that, notwithstanding the neutral gauge boson $Z_{\mu}^{\prime \prime}\equiv Z^{\mu}_3$ does not mix neither with $Z_{\mu}$ nor with $Z_{\mu}^{\prime}$ (in the case $V^\prime \simeq V$ and $v^\prime\simeq v$), it still couples non-universally to ordinary fermions. As a matter of fact, even though the $Z_{\mu}^{\prime \prime}$ couplings are diagonal, its couplings to the third family of quarks are different from the ones to the first two families. Thus, at low energy, we will have tree-level FCNC transmitted by $Z_{\mu }^{\prime \prime}$ which are induced by fermion mixing. This means that in the corresponding low energy effective Lagrangian the chiral $Z_{\mu }^{\prime \prime}$ couplings will in general explicitly depend on the entries of the unitary matrices $V^{\psi}_{L,R}$ that diagonalize the quark Yukawa matrices. So, in the analysis of these FCNC effects a convenient parametrization of the $V^{\psi}_{L,R}$ matrices must be chosen. 

At this point we must notice that if the mass splitting between $Z_2$ and $Z_3$ is so large as the constraints from neutral meson mixing suggest, we must be aware about the fact that the loop contributions coming from the exchange of $Z_2$ can compete with the tree-level effects transmitted by $Z_3$. But, as already discussed, these constraints are rather unreliable, and since the mass matrix in (\ref{remaining}) can be exactly diagonalized we can estimate this mass splitting in the case we are considering, that is $V \simeq V^{\prime} >> v^{\prime}\simeq v$ which, in turn, implies a small mixing. So, neglecting the mixing we have: $m^2_{Z_2} \approx g_4^2 \delta^2 V^2/T^2_W = g_4^2 V^2/(1-T^2_W)$ which, compared with $m^2_{Z_3}=(g^2_4/2)(V^2+v^2)\approx (g^2_4/2) V^2$, shows that in our approximation these masses are of the same order $V$ and, consequently, tree-level FCNC mediated by $Z_3$ will dominate.
%%%%%%%%%%%%%%%%%%%%%%%%%%%%%%%%%%%%%%%%%%%%%%%%%%%%%%%%%%%%

\subsection{The Effective Lagrangian}
We will use the formalism developed in Ref.~\cite{Langacker:2000ju} where general expressions for calculating FCNC effects in models predicting the existence of one extra neutral gauge boson are presented. We start by generalizing this formalism to the case of $N-1$ extra neutral gauge bosons and then we restrict ourselves to the case $N-1=2$.

The Lagrangian for neutral currents in Eq.~(\ref{LNC}) can be rewritten and generalized as
\begin{equation}\label{eq:LNCGEN}
\mathcal L_{\rm NC}=-e A_{\mu}J^{\mu}({\rm EM}) - g_1 Z^{0}_{1,\mu} J^{(1)\,\mu} 
           - \sum\limits_{\alpha=2}^{N-1}g_\alpha Z^{0}_{\alpha,\mu} J^{(\alpha)\,\mu},
\end{equation}
where $Z^0_1\equiv Z$ denotes the neutral gauge boson of the SM and $Z^0_\alpha$ are the new heavy $Z$ bosons (which in general mix with $Z^0_1$).

Following the notation of Ref.~\cite{Langacker:2000ju}, the currents can be written as
\begin{eqnarray}
J^{(m)}_{\mu}&=&\sum\limits_{\psi}\sum\limits_{i,j} \overline{\psi}_i \gamma_{\mu} 
\left[ \epsilon^{\psi(m)}_{{L}_{ij}}P_L 
+ \epsilon^{\psi(m)}_{{R}_{ij}}P_R\right]\psi_j\;,\label{eq:Jmgeneral}
\end{eqnarray}
where the sum extends over all quarks and leptons $\psi_{i,j}$
and $P_{R,L}=(1\pm\gamma_5)/2$. $\epsilon^{\psi(1)}_{{{R,L}}_{ij}}=\epsilon_{R,L}(i) \delta_{ij}$ denotes the SM chiral couplings and $\epsilon^{\psi(m)}_{{{R,L}}_{ij}}$ $(m\neq 1)$ denotes the chiral couplings of the heavy gauge bosons.

After the spontaneous symmetry breaking the physical massive bosons $Z_{\alpha}$ are linear combinations of the gauge eigenstates $Z_{\alpha}^{0}$: 

\begin{equation}
Z_{\alpha}=\sum\limits_{\beta=1}^{N}U_{\alpha\beta}Z_{\beta}^{0}, 
\end{equation} 
where $U$ is an orthogonal $N\times N$ matrix.

The chiral $Z_{\alpha}^{0}$ couplings to fermions in the fermion mass eigenstate basis read

\begin{equation}
E^{\psi(\alpha)}_{L,R}\equiv V^{\psi}_{L,R} \epsilon^{\psi(\alpha)}_{R,L} {V^{\psi}_{L,R}}^{\dagger},\label{eq:EGENERAL}
\end{equation}
where the CKM mixing matrix is given by the combination
\begin{equation}\label{BCKM}
V_{\rm CKM} = V^u_L {V^d_L}^{\dagger}. 
\end{equation}

The four-fermion effective Lagrangian, for the general case of $N$ neutral gauge bosons, reads \cite{Durkin:1985ev}
\begin{equation}
-\mathcal L_{\rm eff}=\frac{4G_F}{\sqrt{2}}\sum\limits_{\alpha=1}^{N}\rho_\alpha \left( \sum\limits_{\beta=1}^{N} U_{\alpha \beta} \frac{g_\beta}{g_1} J_\beta^\mu \right)^{2},\label{eq:LeffLangacker}
\end{equation} 
where $\rho_\alpha \equiv m_W^2/\left(m_\alpha^2cos^2\theta_W\right)$ and $m_\alpha$ is the mass of $Z_ \alpha$.

Using Eqs.~(\ref{eq:Jmgeneral}) and (\ref{eq:EGENERAL}) into (\ref{eq:LeffLangacker}) a general expression for the effective Lagrangian, written in a way useful for further calculations, is obtained as

\begin{eqnarray}
-\mathcal L_{\rm eff}&=&\frac{4G_F}{\sqrt{2}}\sum\limits_{\psi,\chi}\sum\limits_{k,l}\sum\limits_{i,j,m,n}\sum\limits_{X,Y} W^{k l,ijmn}_{X Y} \left( \overline{\psi}_i \gamma_{\mu}P_X \psi_j \right)\left( \overline{\chi}_m \gamma^{\mu}P_Y \chi_n \right),
\label{eq:leffgen}
\end{eqnarray}
where $k,l=1,2,3$; $X$ and $Y$ run over the chiralities $L,R$; $\psi$ and $\chi$ represent classes of fermions with the same SM quantum numbers, i.e. $u,d,e^{-},\nu$, while $i,j,m,n$ are family indexes.
$W^{kl,ijmn}_{P Q}$ is given by

\begin{equation}\label{eq:WIJ}
W^{k l,ijmn}_{X Y} =\frac{g_{k}g_{l}}{g_{1}^{2}}\left(\sum\limits_{r=1}^{N}\rho_r U_{rk}U_{rl}\right)E^{\psi(k)}_{{X}_{ij}}E^{\chi(l)}_{{Y}_{mn}},
\end{equation}
where, in order to have a compact expression for the Lagrangian in Eq.~(\ref{eq:leffgen}), the summation is written so that takes elements of the matrix $U$.

Another way of writing Eq.~(\ref{eq:leffgen}), using the notation of Ref.~\cite{Langacker:2000ju}, is
\begin{eqnarray}\label{Leff}
-\mathcal L_{\rm eff}&=&\frac{4G_F}{\sqrt{2}}\sum\limits_{\psi,\chi}\sum\limits_{i,j,m,n}
\left[
      C^{ij}_{mn} Q^{i j}_{m n}
    + \widetilde{C}^{i j}_{m n} \widetilde{Q}^{i j}_{m n}
    + D^{i j}_{m n} O^{i j}_{m n}
    + \widetilde{D}^{i j}_{m n} \widetilde{O}^{i j}_{m n}
\right]\;,
\end{eqnarray}
with the local operators given by

\begin{eqnarray}\nonumber
Q^{i j}_{m n} &=& \left(\Bar{\psi}_i\gamma^{\mu}P_L\psi_j\right)\left(\Bar{\chi}_m\gamma_{\mu}P_L\chi_n\right),\\ \nonumber
\widetilde{Q}^{ij}_{mn} &=& \left(\Bar{\psi}_i\gamma^{\mu}P_R\psi_j\right)\left(\Bar{\chi}_m\gamma_{\mu}P_R\chi_n\right),\\\nonumber
O^{ij}_{mn} &=& \left(\Bar{\psi}_i\gamma^{\mu}P_L\psi_j\right)\left(\Bar{\chi}_m\gamma_{\mu}P_R\chi_n\right),\\
\widetilde{O}^{ij}_{mn} &=& \left(\Bar{\psi}_i\gamma^{\mu}P_R\psi_j\right)\left(\Bar{\chi}_m\gamma_{\mu}P_L\chi_n\right),
\end{eqnarray}
and the coefficients are

\begin{eqnarray}\nonumber
C^{i j}_{m n}=\sum\limits_{k l} W^{k l,ijmn}_{L L}, &\quad \quad&
\widetilde{C}^{i j}_{m n}=\sum\limits_{k l} W^{k l,ijmn}_{R R},\\
D^{i j}_{m n}=\sum\limits_{k l} W^{k l,ijmn}_{L R}, &\quad \quad&
\widetilde{D}^{i j}_{m n}=\sum\limits_{k l} W^{k l,ijmn}_{R L}. 
\end{eqnarray}

The former expressions are perfectly general for any number of new $Z$ bosons. In the case of only one extra $Z$ boson, from the these expressions and using an orthogonal $2\times 2$ transformation matrix $U$ parametrized by a mixing angle $\theta$, it is straightforward to obtain the formulae for $C^{ij}_{kl}$, $\widetilde{C}^{ij}_{kl}$, $D^{ij}_{kl}$ and $\widetilde{D}^{ij}_{kl}$ in Ref.~\cite{Langacker:2000ju}.  

\subsubsection{The $N=3$ case}
Let us now restrict ourselves to the case $N-1=2$ corresponding to the 3-4-1 extension so that in Eq.~(\ref{eq:LNCGEN}) we have $Z^0_2\equiv Z^\prime$, $Z^0_3\equiv Z^{\prime \prime}$, and a comparison with Eq.~(\ref{LNC}) gives: $g_1 = g_4/C_W$, $g_2=g_X$ and $g_3= g_4/(2\sqrt{2})$. These gauge couplings are expressed in terms of the $SU(2)_L\otimes U(1)_Y$ couplings of the SM by the gauge matching conditions in Eq.~(\ref{matchA}) and, as it is done in studies of low energy FCNC effects associated to extra neutral gauge bosons when only SM particles are present in the effective theory, we neglect their renormalization group (RG) evolution from the high scale $V^\prime \simeq V$ down to the $M_W$ scale [see, for example, Refs.~\cite{Barger:2003hg, Hua:2010we}]. 

Since in our case the $Z-Z^\prime-Z^{\prime \prime}$ mixing occurs between $Z$ and $Z^\prime$ only, the matrix $U$ takes the form 

\begin{equation}\label{eq:Umatrix}
U=
\left(\begin{array}{ccc}
\cos\theta & \sin\theta & 0 \\
-\sin\theta & \cos\theta & 0 \\
0 & 0 & 1
\end{array}\right)\,,
\end{equation}

Now, as already stated, the couplings of $Z_{\mu }^{\prime \prime}$ to the third family of quarks are different from the ones to the first two families, thus allowing FCNC at tree level induced by fermion mixing. These FCNC have consequences on the predictions of $\Delta S=2$ and $\Delta B=2$ observables in the down quark sector as the well measured $\Delta M_K$, $\Delta M_{d/s}$, $\varepsilon_K$ and $\sin \Phi_d$. In order to build the theoretical expressions both for these observables and for the related rare decays we will study let us extract, from Eq.~(\ref{zprimaprima}), the couplings of $Z_{\mu }^{\prime \prime}$ to ordinary down-type quarks. In the gauge eigenstate basis $Q^d=(d_1 d_2 d_3)^T$ the interaction Lagrangian can be written as

\begin{equation}\label{LNCd}
-\mathcal L_{\rm NC}^d = \frac{g_4}{2\sqrt{2}}\sum\limits_{ij=1}^3 \left[\overline{Q}^d_{i} \gamma^\mu (\epsilon^{d(3)}_{Lij} P_L + \epsilon^{d(3)}_{Rij} P_R) Q^d_{j} \right]Z_{\mu }^{\prime \prime},
\end{equation}
where $\epsilon^{d(3)}_{L}= -I_{3\times 3}+2\,\mathrm{diag}(0,0,1)$ and $\epsilon^{d(3)}_{R}=0$. Then, the chiral $Z_3$ couplings, in the mass eigenstate basis ${\mathcal Q}^d=(d\,s\,b)^T$, are
\begin{equation}
E^{d(3)}_{L}=-I_{3\times 3}+2 V^{d}_L\; \mathrm{diag}(0,0,1){V^{d}_L}^{\dagger};\qquad
E^{d(3)}_{R}=0.
 \label{Kij}
\end{equation}

Clearly, FCNC arise in this extension from the non-diagonal elements in the $3\times 3$ matrix $E^{d(3)}_{L}$. That is, the coefficients which determine the contribution of ${\mathcal L}_{\rm eff}$ to $\Delta S = 2$ and $\Delta B =2$ processes are $E^{d(3)}_{{L}_{ij}}$ with $i\neq j$. Therefore, using (\ref{eq:leffgen}) and (\ref{Kij}), the corresponding effective interaction Lagrangian is given by

\begin{eqnarray}
-\mathcal L_{\rm eff} =\frac{4G_F}{\sqrt{2}}&& {\left(\frac{g_{3}}{g_{1}}\right)}^2\rho_3\left(\overline{{\mathcal Q}}^d_{i} \gamma_{\mu}E^{d(3)}_{{L}_{ij}}P_L 
{\mathcal Q}^d_{j} \right)\left( \overline{{\mathcal Q}}^d_{m}\gamma^{\mu}E^{d(3)}_{{L}_{mn}}P_L {\mathcal Q}^d_{n} \right).
\label{LeffFCNC}
\end{eqnarray}
\noindent
The nondiagonal elements of $E^{d(3)}_L$ in Eq.~(\ref{Kij}) read
\begin{equation}
E^{d(3)}_{L\,ij} = 2 V^{d}_{L\, i3} V^{d\ast}_{L\, j3},
\end{equation}
whereas the ratio $(g_{3}/g_{1})$ can be written in terms of the Weinberg angle as
\begin{equation}
\bigg(\frac{g_3}{g_1}\bigg)^2 \ = \ \frac{\cos^2\theta_W}{8}\ .
\end{equation}

As it is well known, in the SM the matrix entries in $V^d_L$ are not observables while the observable quantities only concern the entries in the $V_{\rm CKM}$ matrix. However, from Eq.~(\ref{LeffFCNC}) it is clear that the new physics contributions will be proportional to the entries $V^{d}_{L\, i3}$ in the down quark mixing matrix. We must, then, conveniently parametrize the $V^d_L$ matrix so that we can estimate the size of its entries. This, in turn, requires to determine the number of independent parameters in $V^d_L$. An unitary $3\times 3$ matrix has, in general, nine independent parameters: three real angles and six complex phases. In the CKM mixing matrix five phases can be absorbed in redefinitions of the quark fields, which leaves us with three real angles and one complex phase. In the case of the $V^d_L$ matrix, by analogous redefinitions of the three ordinary down-type quarks, we can absorb three complex phases so that we remain with three real angles and three complex phases. The $V^d_L$ matrix can consequently be conveniently parametrized as the product of three rotations each one of them containing a complex phase \cite{Blanke:2006xr}. We thus get
\begin{widetext}
\begin{equation}
\addtolength{\arraycolsep}{3pt}
V_{L}^d= \begin{pmatrix}
1 & 0 & 0\\
0 & c_{23}^d & s_{23}^d e^{- i\delta^d_{23}}\\
0 & -s_{23}^d e^{i\delta^d_{23}} & c_{23}^d\\
\end{pmatrix}
 \begin{pmatrix}
c_{13}^d & 0 & s_{13}^d e^{- i\delta^d_{13}}\\
0 & 1 & 0\\
-s_{13}^d e^{ i\delta^d_{13}} & 0 & c_{13}^d\\
\end{pmatrix}
 \begin{pmatrix}
c_{12}^d & s_{12}^d e^{- i\delta^d_{12}} & 0\\
-s_{12}^d e^{i\delta^d_{12}} & c_{12}^d & 0\\
0 & 0 & 1\\
\end{pmatrix}
\end{equation}
Performing the product we have
\begin{equation}\label{eq:Vd}
\addtolength{\arraycolsep}{3pt}
V_{L}^d= \begin{pmatrix}
c_{12}^d c_{13}^d & s_{12}^d c_{13}^d e^{-i\delta^d_{12}}& s_{13}^d e^{-i\delta^d_{13}}\\
-s_{12}^d c_{23}^d e^{i\delta^d_{12}}-c_{12}^d s_{23}^ds_{13}^d e^{i(\delta^d_{13}-\delta^d_{23})} &
c_{12}^d c_{23}^d-s_{12}^d s_{23}^ds_{13}^d e^{i(\delta^d_{13}-\delta^d_{12}-\delta^d_{23})} &
s_{23}^dc_{13}^d e^{-i\delta^d_{23}}\\
s_{12}^d s_{23}^d e^{i(\delta^d_{12}+\delta^d_{23})}-c_{12}^d c_{23}^ds_{13}^d e^{i\delta^d_{13}} &
-c_{12}^d s_{23}^d e^{i\delta^d_{23}}-s_{12}^d c_{23}^ds_{13}^d e^{i(\delta^d_{13}-\delta^d_{12})} &
c_{23}^dc_{13}^d\\
\end{pmatrix}
\end{equation}
\end{widetext}
Another way $V_{L}^d$ can be parametrized is \cite{Branco:2004ya}
\begin{equation}
V_L^d \ = \ P \; \tilde V\; K,
\label{vdef}
\end{equation}
where $P = {\rm diag}(e^{i\xi_1},1,e^{i\xi_3})$, $K={\rm
diag}(e^{i\alpha_1},e^{i\alpha_2},e^{i\alpha_3})$ while, using the standard parametrization, the unitary matrix $\tilde V$ can be written in terms of three mixing angles $\theta_{12}$, $\theta_{23}$ and $\theta_{13}$ and one phase $\varphi$ \cite{Nakamura:2010zzi}. Note that this parametrization includes six complex phases but again, by redefinitions of the ordinary down-type quarks, we can absorb three of them. 

It is easy to show that the theoretical expressions resulting from both parametrizations are equivalent via redefinition of phases [See Eqs.~(\ref{phi_1}-\ref{phi_3}) below and Ref.~\cite{Cabarcas:2007my}].

\begin{table}
\caption{\label{tab:2}Values of the experimental and theoretical quantities used as input parameters.}
\begin{center}
\begin{ruledtabular}
\begin{tabular}{llc}
Input & Value & Reference \\ \hline
$G_F$ [GeV] & $1.16637(1)\times 10^{-5}$ & \\
$\alpha$ & $7.2973525376(50)\times 10^{-3}$ & \\
$\alpha_s$ $(m_Z)$ & $0.1184(7)$ & \\
$\sin^2{\theta_W}$ $(m_Z) (\overline{MS})$ & $0.23116(13)$ & \cite{Nakamura:2010zzi} \\ \hline
$|V_{ud}|$ & $0.9728(30)$ & \\
$|V_{us}|$ & $0.2250(27)$ & \\
$|V_{ub}|$ & $0.00427(38)$ & \\
$|V_{cd}|$ & $0.230(11)$ & \\
$|V_{cs}|$ & $0.98(10)$ & \\
$|V_{cb}|$ & $0.0415(7)$ & \\
$|V_{td}|$ & $0.0084(6)$ & \\
$|V_{ts}|$ & $0.0387(21)$ & \\
$|V_{tb}|$ & $0.88(7)$ & \cite{Nakamura:2010zzi}\\ \hline
$m_W$ [GeV] & $80.399(23)$ & \\
$m_Z$ [GeV]  & $91.1876(21)$ & \\
$m_c$ [GeV]  & $1.268(9)$ & \\
$m_t$ [GeV]  & $172.4(1.2)$ & \cite{Nakamura:2010zzi}\\
$\gamma$ & $78(12)$ & \cite{Laiho:2009eu}\\ \hline
$f_K\sqrt{B_K}$ [GeV] & $133(55)\times 10^{-3}$ &  \\
$f_{B_d}\sqrt{B_{B_d}}$ [GeV] & $216(15)\times 10^{-3}$ &  \\
$f_{B_s}\sqrt{B_{B_s}}$ [GeV] & $275(13)\times 10^{-3}$ & \cite{Laiho:2009eu}  \\
$\eta_1$ & $1.32(32)$ & \cite{Buras:1990fn, Brod:2010mj} \\ 
$\eta_2$ & $0.5765(65)$ & \cite{Buras:1990fn, Brod:2010mj}  \\
$\eta_3$ & $0.47(4)$ &  \cite{Brod:2010mj, Herrlich:1996vf} \\
$\eta_B$ & $0.551(7)$ &  \cite{Buras:1990fn, Urban:1997gw} \\
$\kappa_\varepsilon$ & $0.92(1)$ & \cite{Laiho:2009eu} \\ \hline
$m_{K^0}$ [GeV] & $497.614(24)\times 10^{-3}$ &  \\ 
$m_{B_d}$ [GeV] & $5279.50(30)\times 10^{-3}$ &  \\ 
$m_{B_s}$ [GeV] & $5366.3(6)\times 10^{-3}$ &  \\ 
$\Delta m_K$ $[ps^{-1}]$ & $0.5292(9)\times 10^{-2}$ &  \\   
$\Delta m_{B_d}$ $[ps^{-1}]$ & $0.507(5)$ &  \\ 
$\Delta m_{B_s}$ $[ps^{-1}]$ & $17.77(12)$ & \\
$|\varepsilon_K| $ & $2.228(11)\times 10^{-3}$ & \\
$\sin \Phi_d$ & $0.673(23)$ & \cite{Nakamura:2010zzi}\\
\end{tabular}
\end{ruledtabular}
\end{center}
\end{table}

%%%%%%%%%%%%%%%%%%%%%%%%%%%%%%%%%%%%%%%%%%%%%%%%%%%%%%%%%%%%%%%%%%%%%%%%
\section{\label{sec:sec3}$\Delta F=2$ observables} 
We now proceed to build the theoretical expressions for the $\Delta F=2$ ($F=S,B$) neutral meson mixing observables $\Delta M_K$, $\Delta M_{d/s}$, $\varepsilon_K$ and $\sin \Phi_d$ in such a way that the $Z_{\mu }^{\prime \prime}$ contributions enter into the expressions as corrections to the SM predictions, following a similar procedure as for the minimal 3-3-1 model in Ref.~\cite{Promberger:2007py} and for the economical 3-3-1 model in Ref.~\cite{Cabarcas:2007my}. These expressions are functions of the matrix element 
\begin{equation}
M^{P}_{12}\equiv \frac{\langle P^0|{\cal H}_{\rm eff}|\bar P^0\rangle}{2m_P},
\end{equation}
where $P$ stands for $K$, $B_s$ or $B_d$. 
In our case $M^{P}_{12}$ receives both SM contributions arising from standard one loop diagrams and contributions coming from tree-level $Z_3$ exchange, that is 
 
\begin{equation}
M^{P}_{12}=M^{P(\rm SM)}_{12}+M^{P(3-4-1)}_{12}.
\end{equation}
The expressions for the $\Delta F=2$ observables are
\begin{eqnarray}
\Delta m_K & = & 2\,{\rm Re}(M_{12}^{K}) \label{mk}, \\
\Delta m_d & = & 2\,\left|M_{12}^{B_d}\right| \label{md}, \\
\Delta m_s & = & 2\,\left|M_{12}^{B_s}\right| \label{ms}, \\
\varepsilon_K & = & e^{i\phi_\varepsilon}\sin{\phi_\varepsilon}\left(\frac{{\rm Im}(M_{12}^{K})}{\Delta m_K}+P_0 \right), \label{ek} \\
\Phi_d & = & {\rm arg}(M_{12}^{B_d}),  \label{pd} 
\end{eqnarray}
where the term $P_0$ is due to long distance contributions to Kaon mixing. The experimental values for these observables are consigned in Table~\ref{tab:2}.
It is customary to set $\phi_\varepsilon=\pi/4$ and to neglect $P_0$ \cite{Buchalla:1995vs,Buras:2005xt}. For this reason a multiplicative correction factor for $\varepsilon_K$, that accounts for $\phi_\varepsilon \neq \pi/4$ and $P_0\neq 0$, is introduced \cite{Laiho:2009eu,Bu-Gua}. Then, the expression for $\varepsilon_K$ becomes
\begin{eqnarray}
\varepsilon_K & = & \kappa_\varepsilon \exp{\left(i\frac{\pi}{4}\right)}\frac{{\rm Im}(M_{12}^{K})}{\sqrt{2}\Delta m_K}.
\end{eqnarray}
\noindent
The well known SM contributions to $M^{P}_{12}$ are given by \cite{Buchalla:1995vs,Buras:2005xt}

\begin{equation}
M_{12}^{K(\rm SM)} = \frac{G_F^2}{12 \pi^2} f_K^2 B_K m_K m_W^2
\left[ {\lambda_c^*}^2 \eta_1 S_0(x_c) + {\lambda_t^*}^2 \eta_2 S_0(x_t) +
2 {\lambda_c^*} {\lambda_t^*} \eta_3 S_0(x_c, x_t) \right], \label{M12KSM}
\end{equation}
and 
\begin{equation}
M_{12}^{B_q (\rm SM)} = \frac{G_F^2}{12 \pi^2} f_{B_q}^2 B_{B_q} \eta_B m_{B_q} m_W^2 S_0(x_t)  (V_{tq}^\ast V_{tb})^2, \label{M12BSM}
\end{equation}
where $q$ stands for $d$ or $s$. $f_P$ is the $P$-meson decay constant, $B_P$ is the corresponding renormalization scale and scheme invariant bag parameter and the basic electroweak loop contributions, without QCD corrections, are expressed through the functions $S_0(x_i)$ ($x_i=m^2_i/m_W^2$), namely
\begin{equation}\label{s0c}
S_0(x_c)\doteq x_c,
\end{equation}
\begin{equation}\label{s0t}
S_0(x_t)=\frac{4x_t-11x^2_t+x^3_t}{4(1-x_t)^2}-
 \frac{3x^3_t \ln x_t}{2(1-x_t)^3},
\end{equation}
\begin{equation}\label{s0ct}
S_0(x_c, x_t)=x_c\left[\ln\frac{x_t}{x_c}-\frac{3x_t}{4(1-x_t)}-
 \frac{3 x^2_t\ln x_t}{4(1-x_t)^2}\right].
\end{equation}
Renormalization group short-distance QCD effects are described through the renormalization scheme independent factors $\eta_1$, $\eta_2$, $\eta_3$ and $\eta_B$ \cite{Buras:1990fn,Brod:2010mj,Herrlich:1996vf,Urban:1997gw}, and the CKM factors are given by $\lambda_i = V_{is}^* V_{id}^{}$.

Now, from the effective Lagrangian in Eq.~(\ref{LeffFCNC}) the expressions for the 3-4-1 contributions can be obtained. Since only left-left (LL) operators appear in the effective Lagrangian, the QCD RG evolution of the Wilson coefficients $C(\mu)$ after including the $Z_3$ contributions is exactly the same as in the SM. Then, the relevant elements of the evolution matrix $U(\mu, \mu_3)$ from the high scale $\mu_3=m_{Z_3}$ down to the scale $\mu$, in the notation of Ref.~\cite{Buras:2001ra}, can be calculated from

\begin{equation}\label{evol}
[\eta(\mu, \mu_3)]_{\rm VLL}= [\eta(\mu, \mu_t)]_{\rm VLL}[\eta(\mu_t, \mu_3)]_{\rm VLL},
\end{equation}
where $\mu_t={\cal O}(m_t)$.

In general, for $\mu < \mu^\prime$, the correction factor $[\eta(\mu, \mu^\prime)]_{\rm VLL}$ in an effective theory with $f$ active quark flavors is given by
\begin{equation}\label{eta}
[\eta(\mu, \mu^\prime)]_{\rm VLL}= [\eta^{(0)}(\mu)]_{\rm VLL}+\frac{\alpha^{(f)}_s(\mu)}{4\pi}[\eta^{(1)}(\mu)]_{\rm VLL},
\end{equation}
where $[\eta^{(0)}(\mu)]_{\rm VLL}$ and $[\eta^{(1)}(\mu)]_{\rm VLL}$ are the leading order (LO) and the next-to-leading order (NLO) factors, respectively, and in the right-hand side we have suppressed the high scale $\mu^\prime$.

For $K$ and $B$ meson mixing the factor $[\eta(\mu, \mu_t)]_{\rm VLL}$ in Eq.~(\ref{evol}) has been calculated in Ref.~\cite{Buras:2001ra} using $\mu_t=m_t(m_t)=166$ GeV. The results are: $[\eta^K(\mu_L,\mu_t)]_{\rm VLL}=0.788$, for $K^0-\bar{K}^0$ mixing with $\mu_L=2$ GeV the lattice scale where the matching with the lattice results of the associated hadronic matrix elements is done, and $[\eta^B(\mu_b,\mu_t)]_{\rm VLL}=0.842$, for $B^0_{d/s}-\bar{B}^0_{d/s}$ mixing with $\mu_b=4.4$ GeV the mass scale of the bottom quark. In order to calculate $[\eta(\mu_t,\mu_3)]_{\rm VLL}$ we first evaluate $\alpha^{(f)}_s(\mu_3)$ to LO in the $\overline{\rm MS}$ scheme in an effective theory with $f=6$ quark flavors and we impose the continuity relation $\alpha^{(6)}_s(\mu_t)=\alpha^{(5)}_s(\mu_t)$. To this purpose we use

\begin{eqnarray}\nonumber
\frac{1}{\alpha^{(5)}_s(\mu_t)}&=&\frac{1}{\alpha^{(5)}_s(m_Z)}+\frac{\beta^{(5)}_0}{4\pi}\ln \frac{\mu^2_t}{m^2_Z}, \\
\frac{1}{\alpha^{(6)}_s(\mu_3)}&=&\frac{1}{\alpha^{(5)}_s(\mu_t)}+\frac{\beta^{(6)}_0}{4\pi}\ln \frac{\mu^2_3}{\mu^2_t},
\end{eqnarray}
where $\beta^{(f)}_0=11-(2/3)f$ and for $\alpha^{(5)}_s(m_Z)$ we use the central value  $\alpha^{(5)}_s(m_Z)=0.1184$. In this way $[\eta(\mu_t,\mu_3)]_{\rm VLL}$ can be obtained from Eq.~(\ref{eta}) with \cite{Buras:2001ra}

\begin{eqnarray}\nonumber
\left[\eta^{(0)}(\mu_t)\right]_{\rm VLL}&=& \eta^{6/21}_6, \\
\left[\eta^{(1)}(\mu_t)\right]_{\rm VLL}&=& 1.3707(1-\eta_6) \eta^{6/21}_6,
\end{eqnarray}
where $\eta_6 = \alpha^{(6)}_s(\mu_3)/\alpha^{(6)}_s(\mu_t)$. The numerical value of
$[\eta(\mu_t,\mu_3)]_{\rm VLL}$ depends of the value of the high energy scale $\mu_3=m_{Z_3}$. Since in our approximation the $Z_2$ and $Z_3$ mass scales are of the same order (see Sec.~\ref{sec:sec2}), we will select values for $m_{Z_3}$ in the typical range $1-5$ TeV.

With this in mind and defining
 
\begin{eqnarray}\nonumber
U^K_{\rm VLL} &\equiv & [\eta^K(\mu_L,\mu_t)]_{\rm VLL}[\eta(\mu_t,\mu_3)]_{\rm VLL}, \\ U^B_{\rm VLL}  &\equiv & [\eta^B(\mu_b,\mu_t)]_{\rm VLL}[\eta(\mu_t,\mu_3)]_{\rm VLL}, 
\end{eqnarray}
the 3-4-1 contribution to $M^P_{12}$ is given by
\begin{equation}
M_{12}^{P(3-4-1)}=\frac{8\sqrt{2}G_F}{3}U^P_{\rm VLL} \rho_3{\left(\frac{g_{3}}{g_{1}}\right)}^2 m_P f_P^2 \hat B_P \lambda_P^2, \label{M12341}
\end{equation}
where
\begin{eqnarray}\label{bs_1}
\lambda_K & = & s^d_{13}\,s^d_{23}\,c^d_{13}\,e^{i\phi^{''}},
\label{k} \\ \label{bs_2}
\lambda_{B_d} & = & s^d_{13}\,c^d_{23}\,c^d_{13}\,e^{i\phi^{'}},
\label{bd} \\ \label{bs_3}
\lambda_{B_s} & = & s^d_{23}\,c^d_{23}\,(c^d_{13})^2\,e^{-i\phi_3}, \\ \nonumber
\end{eqnarray}
and the phases are defined as
\begin{eqnarray}\label{phi_1}
\phi^{'} & = & -\delta^d_{13},\\ \label{phi_2}
\phi^{''} & = & -(\delta^d_{13}-\delta^d_{23}),\\ \label{phi_3}
\phi_3 & = &\phi^{''} -\phi^{'}. 
\end{eqnarray}

As can be seen from Eq.~(\ref{M12341}), the 3-4-1 contributions $M_{12}^{P(3-4-1)}$ are given in terms of five unknown independent parameters, namely, the mass of the $Z_3$ gauge boson, the angles $\theta_{13}^d$, $\theta_{23}^d$ and the two complex phases $\phi'$ and $\phi''$, where the last four parameters come from the $V_L^d$ mixing matrix.

The $\Delta F=2$ observables discussed in this Section are, in principle, sufficient to set bounds on the relevant parameters of $V_{L}^d$ in Eq.~(\ref{eq:Vd}). 

\begin{figure*}[ht]%Fig.1
\begin{center}
\resizebox{0.98\textwidth}{!}{
\includegraphics[scale=1]{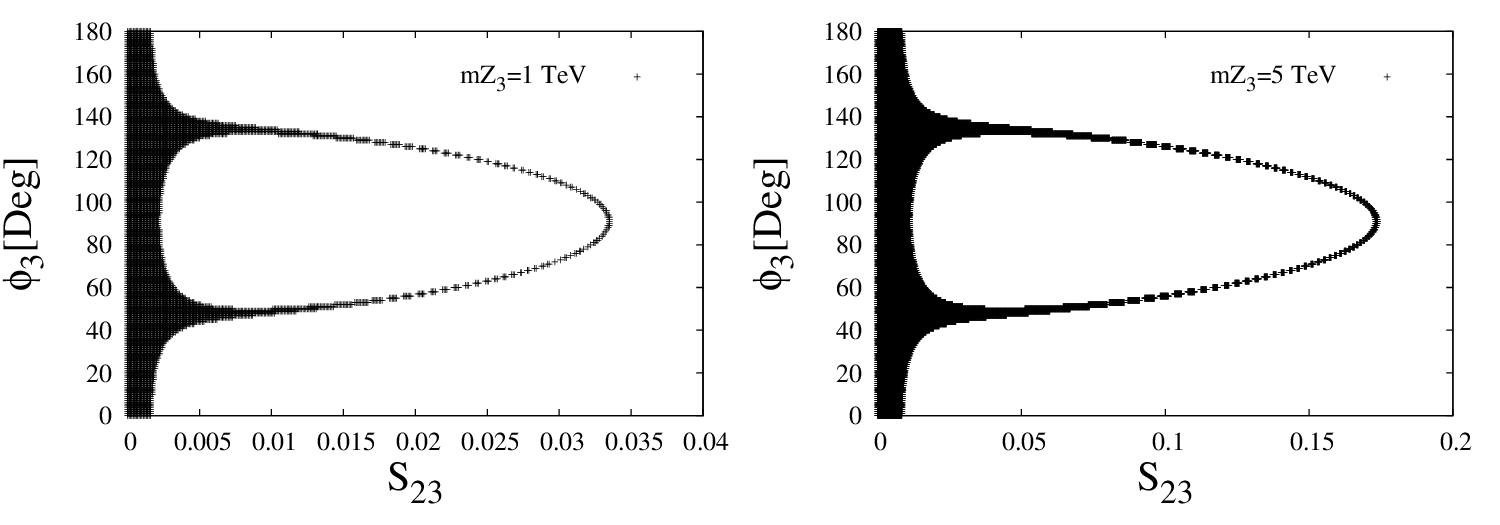} 
}
\caption{\label{fig:1}Allowed $s_{23}^{d}-\phi_3$ region for $s_{13}^{d}=0$, and $m_{Z_3} = ~1 $ TeV (Left), $m_{Z_3} = ~5$ TeV (right).}
\end{center}
\end{figure*}

\begin{figure*} %Fig.2a
\begin{center}
\resizebox{0.98\textwidth}{!}{
\includegraphics[scale=1]{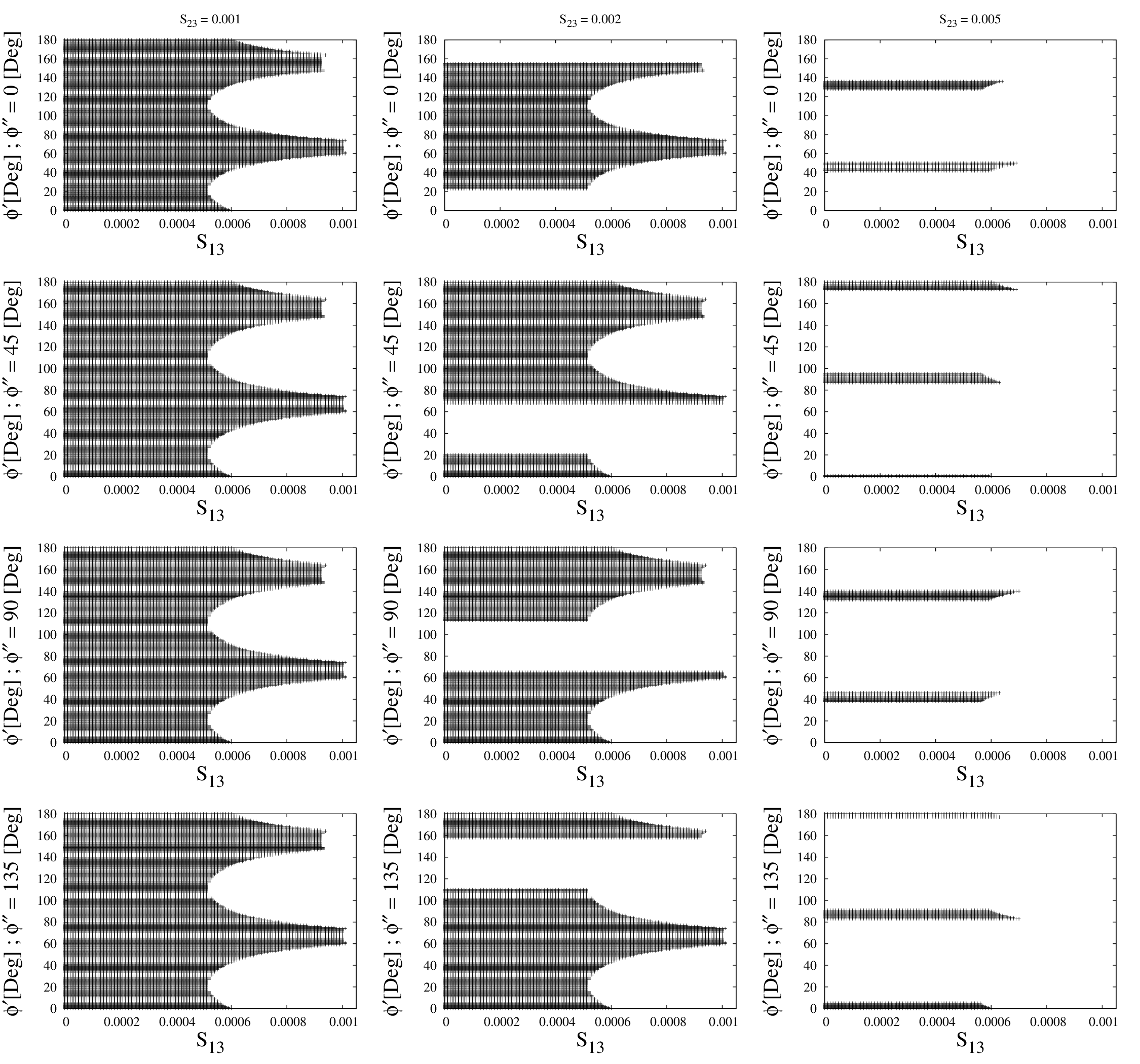} 
}
\caption{\label{fig:2a}Allowed $s_{13}^{d}-\phi'$ region for $m_{Z_3}=1$ TeV and for different values of $\phi''$ and increasing values of $s_{23}^{d}$: $0.001$ (left), $0.002$ (center) and $0.005$ (right).}
\end{center}
\end{figure*}

\begin{figure*}[ht] %Fig.3
\begin{center}
\resizebox{0.98\textwidth}{!}{
\includegraphics[scale=1]{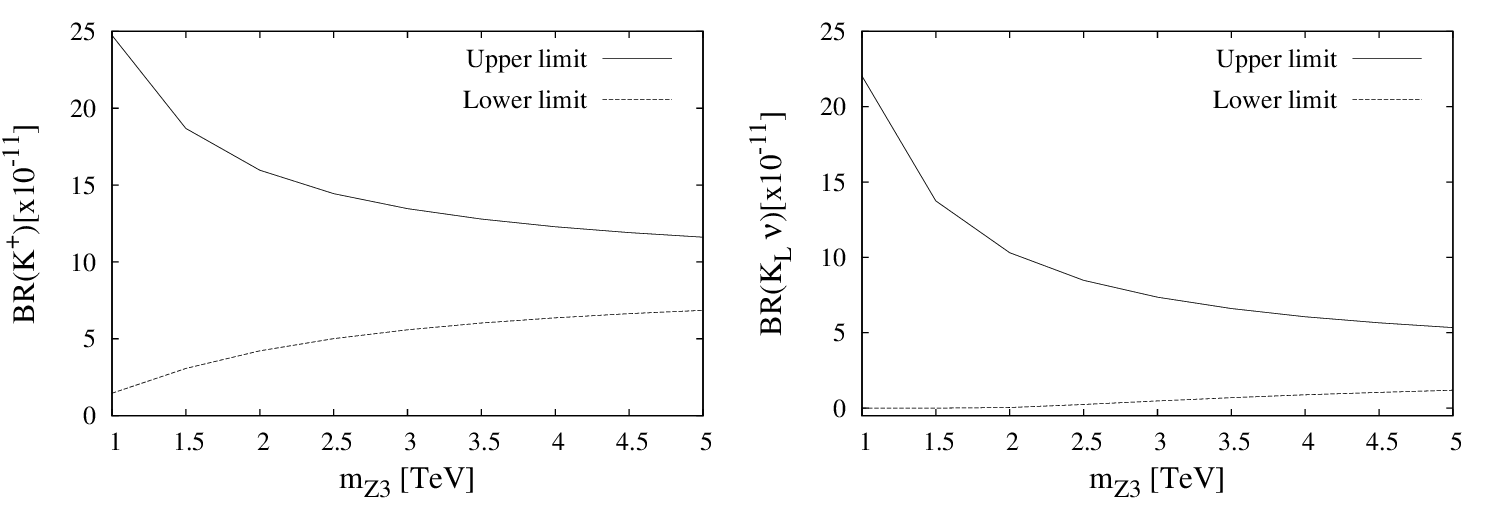} 
}
\caption{\label{fig:3}Upper and lower bounds on the BR for the decay $K^+ \to \pi^+ \bar \nu \nu$ (left), and for the decay $K_{L} \to \pi^0\nu\bar\nu$ (right).}
\end{center}
\end{figure*}

\begin{figure*}[ht] %Fig.3b
\begin{center}
\resizebox{0.98\textwidth}{!}{
\includegraphics[scale=1]{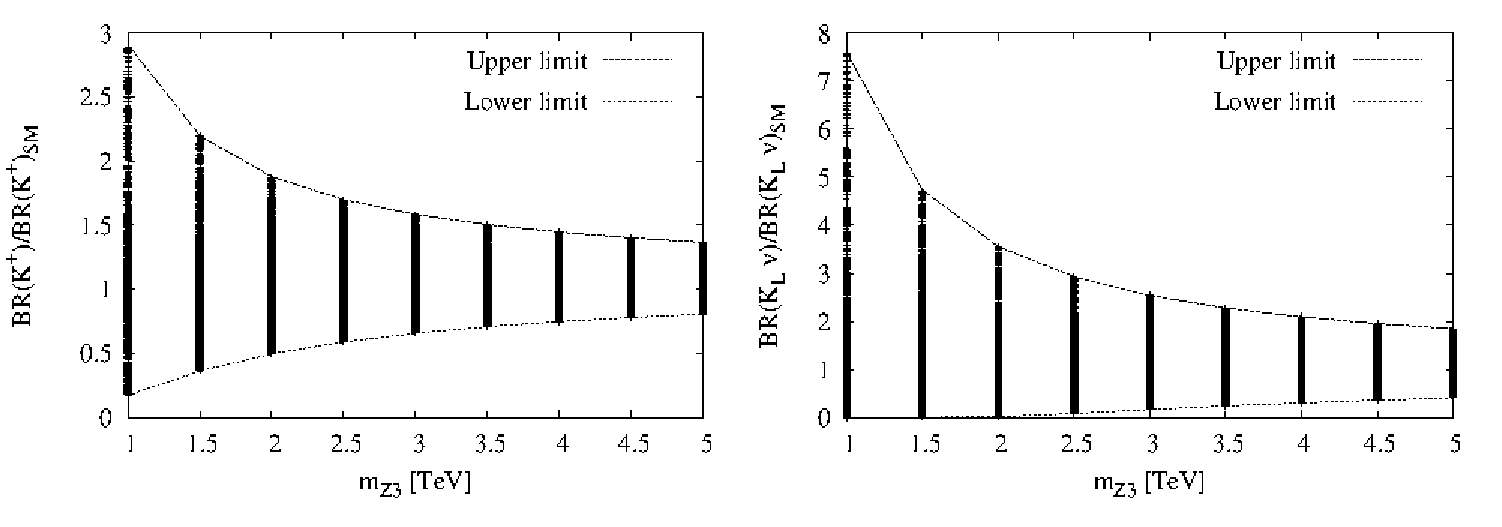} 
}
\caption{\label{fig:3b}${\mbox{BR}(K^+\rightarrow\pi^+\nu\bar\nu)}/{{\mbox{BR}(K^+\rightarrow\pi^+\nu\bar\nu)}_{\rm SM}}$ (left) and ${\mbox{BR}(K_{ L}\rightarrow\pi^0\nu\bar\nu)}/{{\mbox{BR}(K_{ L}\rightarrow\pi^0\nu\bar\nu)}_{\rm SM}}$ (right). Upper and lower bounds for different values of $m_{Z_3}$ are shown.}
\end{center}
\end{figure*}

\begin{table}[ht]
\caption{\label{tab:3}Upper limit for $\theta_{13}$ and $\theta_{23}$ for different values of $m_{Z_3}$.}
\begin{center}
\begin{ruledtabular}
\begin{tabular}{ccc}
$m_{Z_3}$& $\theta_{13,{\rm max}}$ & $\theta_{23,{\rm max}}$ \\ \hline
1~TeV & $8.73\times 10^{-4}$ & $3.14\times 10^{-2}$ \\ \hline
3~TeV &$2.97\times 10^{-3}$ & $9.42\times 10^{-2}$ \\ \hline
5~TeV & $5.06\times 10^{-3}$ & $1.75\times 10^{-1}$ \\
\end{tabular}
\end{ruledtabular}
\end{center}
\end{table}

%%%%%%%%%%%%%%%%%%%%%%%%%%%%%%%%%%%%%%%%%%%%%%%%%%%%%%%%%%%%%%%%%%%%%%%%%%
\section{\label{sec:sec4}Theoretical expressions for Rare Decay Amplitudes}
In the next Section, using the observables related to meson mixing, the bounds on the parameter space associated to the relevant entries of the down quark mixing matrix will be established. Our next task will be to study the implications of these bounds on several rare decay amplitudes following a similar approach as in Ref.~\cite{Promberger:2007py}. The decays we will study are the theoretically very clean rare decays $K^+ \to \pi^+  \nu \bar \nu$, $K_L \to \pi^0 \nu \bar \nu$, and the clean rare decays $B_{d/s} \to \mu^+ \mu^-$, $K_L \to \pi^0 l^+ l^-$ ($l=e,\, \mu$). 
 
As it is known, the rare decays we are interested in are governed both by electroweak- and photon-penguins and by leptonic box diagram contributions. In the SM these contributions are described, in the LO, by the corresponding Inami-Lim functions $C_0(x_t)$, $D_0(x_t)$ and $B_0(x_t)$ which, in the expressions for the decay amplitudes, always appear in the gauge invariant combinations: $X_0(x_t) = C_0(x_t)-4B_0(x_t)$, $Y_0(x_t) = C_0(x_t)-B_0(x_t)$, $Z_0(x_t) = C_0(x_t)+(1/4)D_0(x_t)$ \cite{Buchalla:1990qz}. 
In the NLO these combinations will be written as $X(x_t)$, $Y(x_t)$ and $Z(x_t)$. Since the $Z_3$ contribution to the effective Hamiltonians for the rare decays we are considering have the same operator structure than the ones in the SM, that is they have a $(V-A)(V-A)$ form, its effects on the various decays can be encoded by appropriate redefinitions of the $X(x_t)$, $Y(x_t)$ and $Z(x_t)$ functions in the form  

\begin{eqnarray}\nonumber
X^\prime(x_t)&=&X^{(SM)}(x_t)+\Delta X\,, \\ \nonumber
Y^\prime(x_t)&=&Y^{(SM)}(x_t)+\Delta Y\, \\
Z^\prime(x_t)&=&Z^{(SM)}(x_t)+\Delta Z\,.
\end{eqnarray}
As we will see, in the 3-4-1 extension the redefined functions are in general complex.  We will write them as they appear in Ref.~\cite{Buchalla:1995vs} so that the modifications lead to the correct results in the 3-4-1 extension including the corresponding imaginary part. For the decays $K_L \to \pi^0 l^+ l^-$, in which there is also a right-handed contribution, the $Z_3$ effects will be absorbed into the matching conditions of the Wilson coefficients.

\subsection{ $K\to \pi \nu \bar \nu$}
For $K\to \pi \nu \bar \nu$ there exists a charged decay $K^+\rightarrow\pi^+\nu\bar\nu$, and a neutral one $K_{ L}\rightarrow\pi^0\nu\bar\nu$ \cite{Buras:2004uu}. For both of them the NLO effective Hamiltonian is given by
\begin{equation}
H^{\rm SM}_{\rm eff}=\frac{G_F}{\sqrt 2}\frac{\alpha}{2\pi 
\sin^2\theta_W}
 \sum_{l=e,\mu,\tau}\left( V^{\ast}_{cs}V_{cd} X^l_{\rm NL}+
V^{\ast}_{ts}V_{td} X(x_t)\right)
 (\bar sd)_{V-A}(\bar\nu_l\nu_l)_{V-A} \,, \\
\end{equation}
where the functions $X_{\rm NL}$ and $X(x_t)$ represent the charm and top-loop contributions, respectively.
\noindent
Summing over the three neutrino flavors and collecting the charm contributions in $P_c(X)=0.41\pm0.05$ \cite{Bu-Brod}, the BR for $K^+\rightarrow\pi^+\nu\bar\nu$ can be expressed as
\begin{widetext}
\begin{eqnarray}\label{kpvv}
\mbox{BR}(K^+\to\pi^+\nu\bar\nu) &=&\kappa_+\cdot\left[\left({\rm Im}\left(\frac{\lambda_t}{\lambda^5}X(x_t)\right)\right)^2 + 
\left({\rm Re}\left(\frac{\lambda_c}{\lambda}P_c(X)\right)\right.\right. \\ \nonumber
& &+\left.\left.{\rm Re}\left(\frac{\lambda_t}{\lambda^5}X(x_t)\right)\right)^2\right],
\end{eqnarray}
\end{widetext}
where \cite{Isidori:2005xm}
\begin{equation}\label{kpovv}
\kappa_+=r_{K^+}\frac{3\alpha^2 \mbox{BR}(K^+\to\pi^0 e^+\nu)}{
 2\pi^2\sin^4\theta_W}\lambda^8=(5.26\pm 0.06)\cdot 10^{-11}
\left[\frac{\lambda}{0.225}\right]^8.
\end{equation}
For $K_{L}\rightarrow\pi^0\nu\bar\nu$ it is found
\begin{equation}\label{bklpn}
\mbox{BR}(K_L\to\pi^0\nu\bar\nu)=\kappa_L\cdot
\left({\rm Im} \left(\frac{\lambda_t}{\lambda^5}X(x_t)\right)\right)^2,
\end{equation}
with
\begin{equation}\label{kapl2}
\kappa_L=\kappa_+ \frac{r_{K_L}}{r_{K^+}}
\frac{\tau(K_L)}{\tau(K^+)}=
(2.29\pm 0.03)\cdot 10^{-10}\left[\frac{\lambda}{0.225}\right]^8,
\end{equation}
where $\lambda$ is a parameter of the Wolfenstein parametrization of the CKM matrix \cite{Wolfenstein:1983yz} and is set equal to $s_{12}$ of the standard parametrization \cite{Nakamura:2010zzi}. 

For both decays the contribution from the 3-4-1 extension to the effective Hamiltonian comes from a tree diagram transmitted by the $Z''$ boson and has the form
\begin{equation}
H_{\rm eff}^{Z''}= \sum_{l=e,\mu,\tau} \frac{2 G_F}{\sqrt{2}} {\left(\frac{g_{3}}{g_{1}}\right)}^2\rho_3 V^{d}_{L\, 23} V^{d\ast}_{L\, 13}  (\bar s d)_{V-A}(\bar \nu_l  \nu_l)_{V-A}\,.
\end{equation} 
This contribution can be included as a modification $\Delta X$ of the function $X(x_t)$ which reads
\begin{equation}\label{X341}
\Delta X_{K \pi \nu \nu}=\frac{4\pi\sin^2{\theta_W}}{\alpha} {\left(\frac{g_{3}}{g_{1}}\right)}^2\rho_3 \frac{V^{d}_{L\, 23} V^{d\ast}_{L\, 13} }{V_{ts}^* V_{td}}\,.
\end{equation}

\subsection{$B_{d/s} \to \mu^+ \mu^-$}
These decays are entirely determined by the top contribution so that the SM effective Hamiltonian in the NLO is given by
\begin{equation}
H_{\rm eff}^{B_{d/s}\mu \mu}= -\frac{G_F}{\sqrt{2}}\frac{\alpha}{2 \pi s_W^2} (V^*_{tb}V_{td/s}) Y(x_t) (\bar b q)_{V-A}(\bar \mu \mu)_{V-A}\,.
\end{equation}
From here, the expressions for the branching fractions are obtained as
\begin{equation}\label{blls}
\mbox{BR}(B_q\to \mu^+\mu^-)=
\tau_{B_q} \frac{G_F^2}{\pi} m_{B_q}
\left(\frac{\alpha f_{B_q} m_{\mu}}{4 \pi \sin^2 \theta_W} \right)^2 \sqrt{1-4 \frac{m_{\mu}^2}{m_{B_q}^2}}
|V^\ast_{tb}V_{tq} Y(x_t)|^2.
\end{equation}

The 3-4-1 contribution to $B_{d/s} \to \mu^+ \mu^-$ is found to be
\begin{equation}
H_{\rm eff}^{Z''}=\frac{2 G_F}{\sqrt{2}} {\left(\frac{g_{3}}{g_{1}}\right)}^2\rho_3 V^{d}_{L\, 33} V^{d\ast}_{L\, 13/23} (\bar b q)_{V-A}(\bar \mu \mu)_{V-A},
\end{equation}
which we now absorb into a modification of $Y(x_t)$ as 
\begin{equation}\label{Y341}
\Delta Y_{B\mu \mu}=-\frac{4\pi\sin^2{\theta_W}}{\alpha} {\left(\frac{g_{3}}{g_{1}}\right)}^2\rho_3 \frac{V^{d}_{L\, 33} V^{d\ast}_{L\, 13/23} }{V_{tb}^* V_{td/ts}}\,.
\end{equation}

\subsection{$K_L \to \pi^0 l^+ l^-$}
In the SM the CP-violating part of the effective Hamiltonian, after neglect QCD penguin operators, is given by
\begin{equation}
H_{\rm eff}^{K\pi ll}=-\frac{G_F}{\sqrt{2}} V_{ts}^* V_{td}(y_{7V} Q_{7V}+y_{7A} Q_{7A}) \,, 
\end{equation}
where  $Q_{7V}=(\bar s d)_{V-A} \bar e\gamma^{\mu}e$ and $Q_{7A}=(\bar sd)_{V-A}\bar e\gamma^{\mu} \gamma^5e$ are the vector- and axial-vector operators which contribute and that originate from $\gamma$- and $Z$-penguin and box diagrams. The matching conditions of the Wilson coefficients $y_{7V}$ and $y_{7A}$ are given by

\begin{equation}
y_{7V}=\frac{\alpha}{2 \pi} \left( \frac{Y_0(x_t)}{\sin^2{\theta_W}} -4 Z_0(x_t)+P_0\right), 
\end{equation}
and
\begin{equation}
y_{7A}=-\frac{\alpha}{2 \pi} \frac{Y_0(x_t)}{\sin^2{\theta_W}}.
\end{equation}
In $y_{7V}$ a small term $P_E$ has been neglected and we use the normalization $P_0=2.89\pm 0.06$ \cite{Buchalla:2003sj,Isidori:2004rb}.
 
For these decays the contributions from new physics lead to the effective Hamiltonian
\begin{equation}
H_{\rm eff}^{Z''}=\frac{2 G_F}{\sqrt{2}} {\left(\frac{g_{3}}{g_{1}}\right)}^2\rho_3 V^{d}_{L\, 23} V^{d\ast}_{L\, 13}\left(Q_{7V}-Q_{7A} \right). 
\end{equation}

Following \cite{Promberger:2007py}, instead of absorbing these new contributions into modifications of the Inami-Lim functions, they will be absorbed into the matching conditions of the Wilson coefficients in the form
\begin{eqnarray}
\Delta y_{7V} &=& -2{\left(\frac{g_{3}}{g_{1}}\right)}^2\rho_3 \frac{(V^{d}_{L\, 23} V^{d\ast}_{L\, 13})}{V_{ts}^* V_{td}} \label{y_V}, \\
\Delta y_{7A} &=& 2{\left(\frac{g_{3}}{g_{1}}\right)}^2\rho_3 \frac{(V^{d}_{L\, 23} V^{d\ast}_{L\, 13})}{V_{ts}^* V_{td}}\label{y_A}.
\end{eqnarray}

The expressions for the BR, including the long-distance indirectly CP-violating terms and their interference with the short-distance contributions, are \cite{Mescia:2006jd}  

\begin{equation}\label{BrKpiLL}
Br(K_L\to\pi^0\ell^+\ell^-)=\left(C_\text{dir}^\ell\pm
  C_\text{int}^\ell\left|a_s\right| +
  C_\text{mix}^\ell\left|a_s\right|^2+C_\text{CPC}^\ell\right)\cdot
10^{-12}\,,
\end{equation}
where
\begin{align}
&C_\text{dir}^e = (4.62\pm0.24)(\omega_{7V}^2+\omega_{7A}^2)\,,&\qquad&
C_\text{dir}^\mu =(1.09\pm0.05)(\omega_{7V}^2+2.32\omega_{7A}^2)\,,\\
&C_\text{int}^e = (11.3\pm0.3)\omega_{7V}\,,&\qquad&
C_\text{int}^\mu = (2.63\pm0.06)\omega_{7V}\,,\\
&C_\text{mix}^e = 14.5\pm0.05\,,&\qquad&
C_\text{mix}^\mu = 3.36\pm0.20\,,\\
&C_\text{CPC}^e \simeq 0\,,&\qquad&
C_\text{CPC}^\mu = 5.2\pm1.6\,,\\
&&&\hspace{-1.5cm}\left|a_s\right|=1.2\pm0.2,
\end{align}
with $w_{7A,7V}=\operatorname{Im}\left(\lambda_{t}y_{7A,7V}\right)
/\operatorname{Im}\lambda_{t}$.

\begin{figure}[ht] %Fig.3c
\begin{center}
\resizebox{0.55\textwidth}{!}{
\includegraphics[scale=0.7]{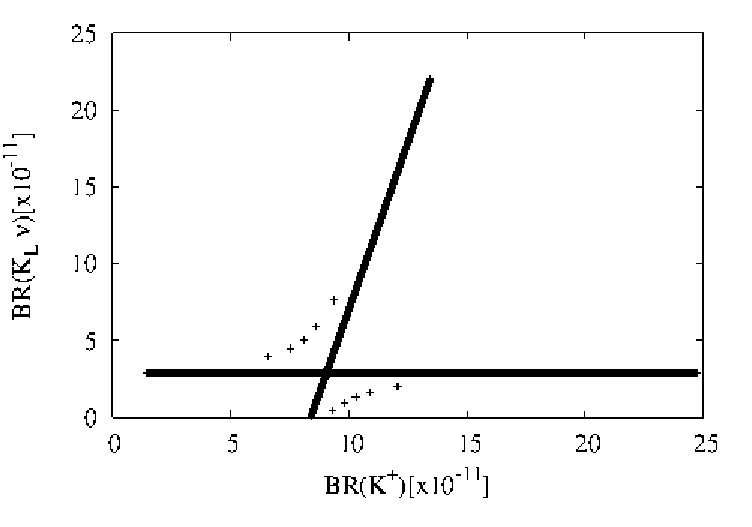} 
}
\caption{\label{fig:3c}$\mbox{BR}(K_{L} \to \pi^0\nu\bar\nu)-\mbox{BR}(K^+ \to \pi^+ \bar \nu \nu)$ plane for $m_{Z_3}= 1$ TeV.}
\end{center}
\end{figure}
 
\begin{figure*}[ht] %Fig.4
\begin{center}
\resizebox{0.98\textwidth}{!}{
\includegraphics[scale=1]{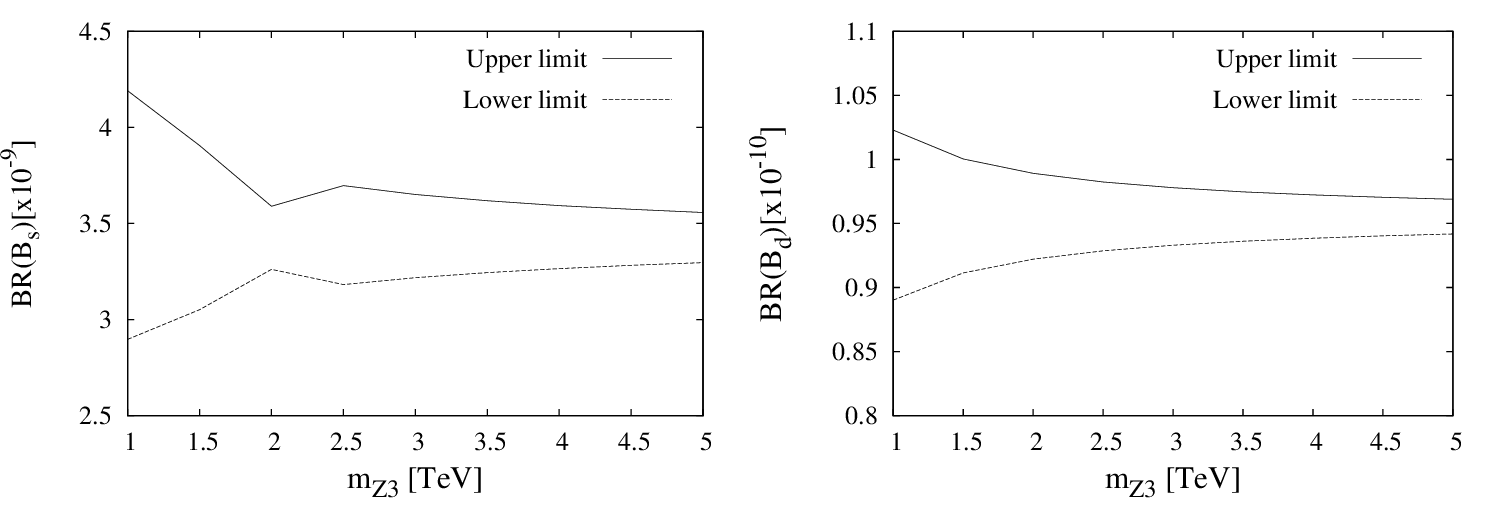} 
}
\caption{\label{fig:4}Upper and lower bound on the BR for the decay $B_s\to \mu^+\mu^-$ (left), and for the decay $B_d\to \mu^+\mu^-$ (right) as a function of $m_{Z_3}$.}
\end{center}
\end{figure*}

\begin{figure*} %Fig.4b
\begin{center}
\resizebox{0.98\textwidth}{!}{
\includegraphics[scale=1]{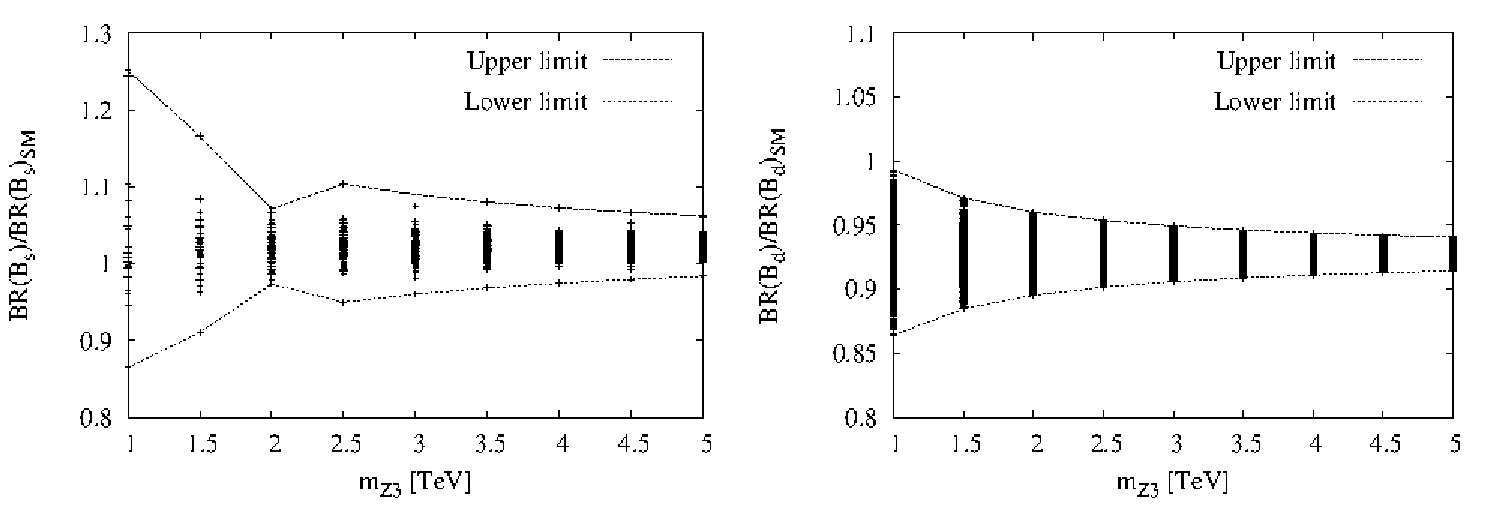} 
}
\caption{\label{fig:4b}${\mbox{BR}(B_s\to \mu^+\mu^-)}/{{\mbox{BR}(B_s\to \mu^+\mu^-)}_{\rm SM}}$ (left) and ${\mbox{BR}(B_d\to \mu^+\mu^-)}/{{\mbox{BR}(B_d\to \mu^+\mu^-)}_{\rm SM}}$ (right) for different values of $m_{Z_3}$, including upper and lower bounds.}
\end{center}
\end{figure*}

%%%%%%%%%%%%%%%%%%%%%%%%%%%%%%%%%%%%%%%%%%%%%%%%%%%%%%%%%%%%%%%%%%%%%%%%%%%
\section{\label{sec:sec5}Constraints and implications for rare decays}
In this Section we numerically evaluate the theoretical expressions obtained in the two previous Sections. As noted above, the 3-4-1 contributions $M_{12}^{P(3-4-1)}$ to the matrix elements in Eqs.~(\ref{mk}-\ref{pd}) are given in terms of the five unknown independent parameters $m_{Z_3}$, the angles $\theta_{13}^d$, $\theta_{23}^d$ and two complex phases, $\phi'$ and $\phi''$. We perform two related numerical analysis: in the first one, we use the well-measured observables $\Delta M_K$, $\Delta M_{d/s}$, $\varepsilon_K$ and $\sin \Phi_d$ to constrain the angles and phases appearing in the $V_L^d$ mixing matrix for selected values of the $Z_3$ mass.  In the second analysis we study the implications of these bounds on the rare decays previously mentioned. In this context we are mainly interested in obtaining upper and lower bounds for these decays as a function of the $Z_3$ mass in order to confront them with the present experimental data and with the SM predictions.
The updated experimental data for the input parameters are also given in Table~\ref{tab:2}.

\subsection{Constraints from $K^0 - \bar K^0$ and $B^0_q - \bar B^0_q$ mixing}
We start by focusing on the bounds on the parameter space of the 3-4-1 extension  associated to the down-like quark mixing parameters. 
Two possible analyses can be carried out. The first one, followed in several works both in 3-3-1 models \cite{331-2} and in 3-4-1 extensions \cite{341-2, Nisperuza:2009xm, Villada:2009iu}, consist in to assume a texture for the mixing matrix in order to obtain bounds on the mass of the heavy gauge bosons. In the opposite approach, one can fix these masses and obtain some information on the size of the corresponding mixing matrix elements \cite{Liu:1993gy, Cabarcas:2007my, Promberger:2007py}. Here we follow this second approach with fixed values of $m_{Z_3}$ in the range $1-5$ TeV.

In previous studies on flavour physics observables in different models, the numerical analysis was simplified by setting all input parameters to their central values and allowing instead $\Delta M_K$, $\varepsilon_K$, $\Delta M_d$, $\Delta M_s$, $|\varepsilon_K|$ and $\sin \Phi_d$ to differ from their experimental values by $\pm 50\%$, $\pm 40\%$, $\pm 40\%$, $\pm 40\%$, $\pm 20\%$ and $\pm 8\%$, respectively.  
This simplifying assumption was justified in order to determine the size of possible effects on observables that have not been well measured so far. In recent analysis, however, an improved error analysis has been done that enable us to draw more accurate conclusions in view of the recent significant improvements in the experimental constraints and in the lattice determinations of the non-perturbative parameter. Therefore, in what follows, we will take all input parameters to be flatly distributed within their $1\sigma$ ranges indicated in Table~\ref{tab:2}. At the same time, we require the observables $|\varepsilon_K|$, $\Delta M_d$, $\Delta M_s$ and $\sin \Phi_d$, resulting from the SM and the 3-4-1 contributions, to lie within their experimental $1\sigma$ ranges. In the case of $\Delta M_K$, where the theoretical uncertainty is large due to unknown long-distance contributions, we allow the generated value to lie within $\pm 30\%$ of its experimental central value.

In dealing with the values of the CKM matrix parameters quoted in Table~\ref{tab:2}, instead of using the full CKM angle fits, we take only the values related to direct experimental measures with the largest uncertainties taken into account. In this way we avoid possible modifications from the 3-4-1 contributions to the one-loop SM processes included in the global fits and leave the largest room available for possible new physics effects, respecting the well measured values. For the CP-violating parameter $\gamma$ we take its direct determination from the model-independent UTfit analysis of $B \to D^{(*)}K^{(*)}$ decays.

We then proceed to scan the parameter space fixing $m_{Z_3}$ and generating a large number of points that we call an ``effective" parameter space. For each fixed value of the $Z_3$ mass, this space consists of more than $1\times 10^{6}$ points that fulfil the requirements listed above, where all angles are varied in the interval $[0,\pi/2]$, all phases between 0 and 2$\pi$, and all input parameters are varied in their $1\sigma$ ranges.

With these data we can plot contours setting bounds on some of the parameters. For example, let us take the mixing angle $\theta_{13}^{d}=0$. In this case, Eq.~(\ref{bs_3}) determines the allowed region in the $s_{23}^{d}-\phi_3$ plane, which is shown in Fig.~\ref{fig:1} for $m_{Z_3}=1$ TeV and $m_{Z_3}=5$ TeV. 
These plots allows us to set upper bounds on $\vert s^d_{23}\vert$ which depend on the value of $m_{Z_3}$. It is seen that the upper bound for $s^d_{23}$ increase for increasing values of $m_{Z_3}$. Fig.~\ref{fig:1} also shows that, at the $1\sigma$ confidence level, there is a large and typical region which is excluded and that prefers values for $\phi_3$ around $48^\circ$ and $136^\circ$ independently of the $Z_3$ mass.

Now we explore, for several values of $m_{Z_3}$, the allowed regions in the plane $s^d_{13}-\phi^\prime$ using non-zero values of $s^d_{23}$ and selected values of $\phi''$. This is shown in Fig.~\ref{fig:2a} for $m_{Z_3}=1$. We have taken values for $s^d_{23}$ consistent with the ones in the allowed $s_{23}^{d}-\phi_3$ plane and we have fixed $\phi''$ at the values $0^\circ$, $45^\circ$, $90^\circ$ and $135^\circ$. This figure shows that for increasing values of $s^d_{23}$ and depending of the $\phi''$ value, the allowed region reduces considerably involving both the selection of very specific values of $\phi^\prime$ and a decrease in the upper bound for $s^d_{13}$.

With the data from the effective parameter space we can estimate the order of magnitude of the upper limits for $\theta_{13}^{d}$ and $\theta_{23}^{d}$ for several values of the $Z_3$ mass. The results are collected in Table~\ref{tab:3} which shows that these upper bounds increase for increasing values of $m_{Z_3}$. This allow us to elucidate the size of the allowed region in the parameter space. 
 
\begin{table*}[t]
\caption{\label{tab:4}Present experimental data and SM predictions for the rare decays considered in this work.}
\begin{center}
\begin{ruledtabular}
\begin{tabular}{cccc}
Label & Decay & Experimental & SM \\ \hline
$\mbox{BR}(K^+)$ & $\mbox{BR}(K^+\rightarrow\pi^+\nu\bar\nu)$ & $(1.7 \pm 1.1)\cdot 10^{^{-10}}$ & $(7.81^{+0.80}_{-0.71}\pm0.29)\cdot 10^{^{-11}}$\\
$\mbox{BR}(K_L \nu)$ & $\mbox{BR}(K_{ L}\rightarrow\pi^0\nu\bar\nu)$ & $<6.7 \cdot 10^{-8} \quad (90\% \mbox{CL})$ & $(2.43^{+0.40}_{-0.37} \pm 0.06) \cdot 10^{-11}$\\
$\mbox{BR}(B_s)$ & $\mbox{BR}(B_s\to \mu^+\mu^-)$ & $< 1\cdot 10^{-7} \quad (90\% \mbox{CL})$ & $(3.35\pm0.32)\cdot 10^{-9}$\\
$\mbox{BR}(B_d)$ & $\mbox{BR}(B_d\to \mu^+\mu^-)$ & $<3 \cdot 10^{-8} \quad (90\% \mbox{CL})$ & $(1.03 \pm 0.09) \cdot 10^{-10} $\\
$\mbox{BR}(K_L e)$ & $\mbox{BR}(K_L \to \pi^0 e^+ e^-)$ &$< 28 \cdot 10^{-11} \quad (90\% \mbox{CL})$& $(3.54^{+0.98}_{-0.85}) \cdot 10^{-11}$\\
$\mbox{BR}(K_L \mu)$ & $\mbox{BR}(K_L \to \pi^0 \mu^+ \mu^-)$ &$< 38 \cdot 10^{-11}  \quad (90\% \mbox{CL})$& $(1.41^{+0.28}_{-0.26}) \cdot 10^{-11}$\\
\end{tabular}
\end{ruledtabular}
\end{center}
\end{table*}

\begin{figure*}[ht] %Fig.5
\begin{center}
\resizebox{0.98\textwidth}{!}{
\includegraphics[scale=1]{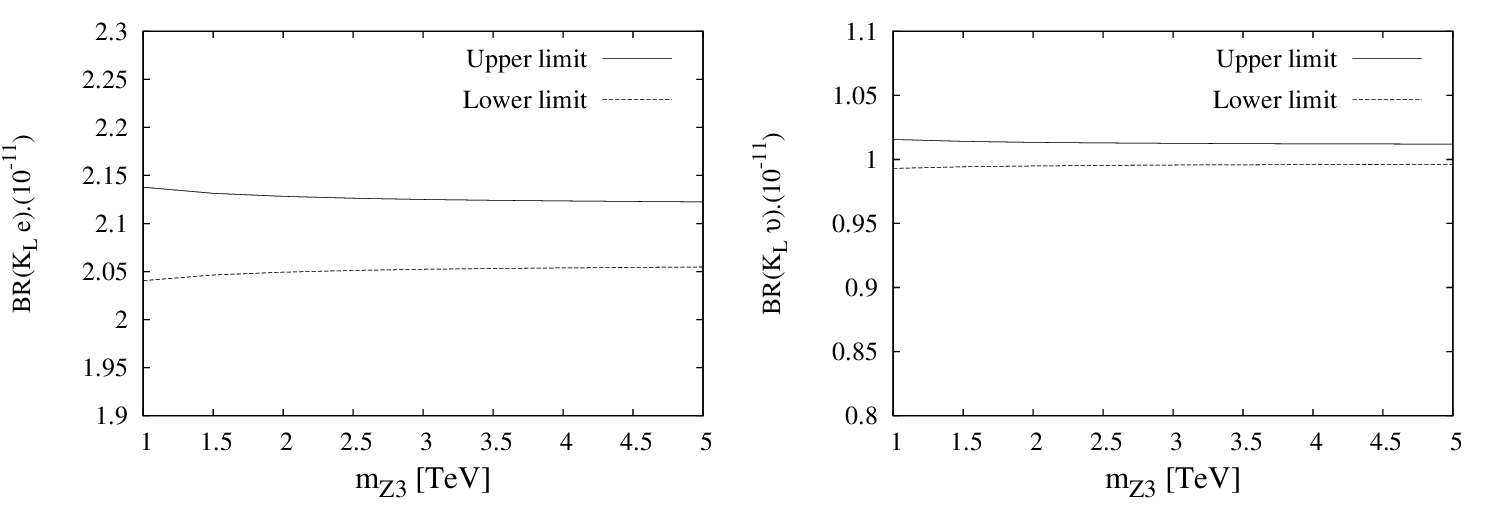} 
}
\caption{\label{fig:5}Upper and lower bound on the BR for the decay $K_L \to \pi^0 e^+ e^-$ (left), and for the decay $K_L \to \pi^0 \mu^+ \mu^-$ (right).}
\end{center}
\end{figure*} 

\begin{figure*} %Fig.5b
\begin{center}
\resizebox{0.98\textwidth}{!}{
\includegraphics[scale=1]{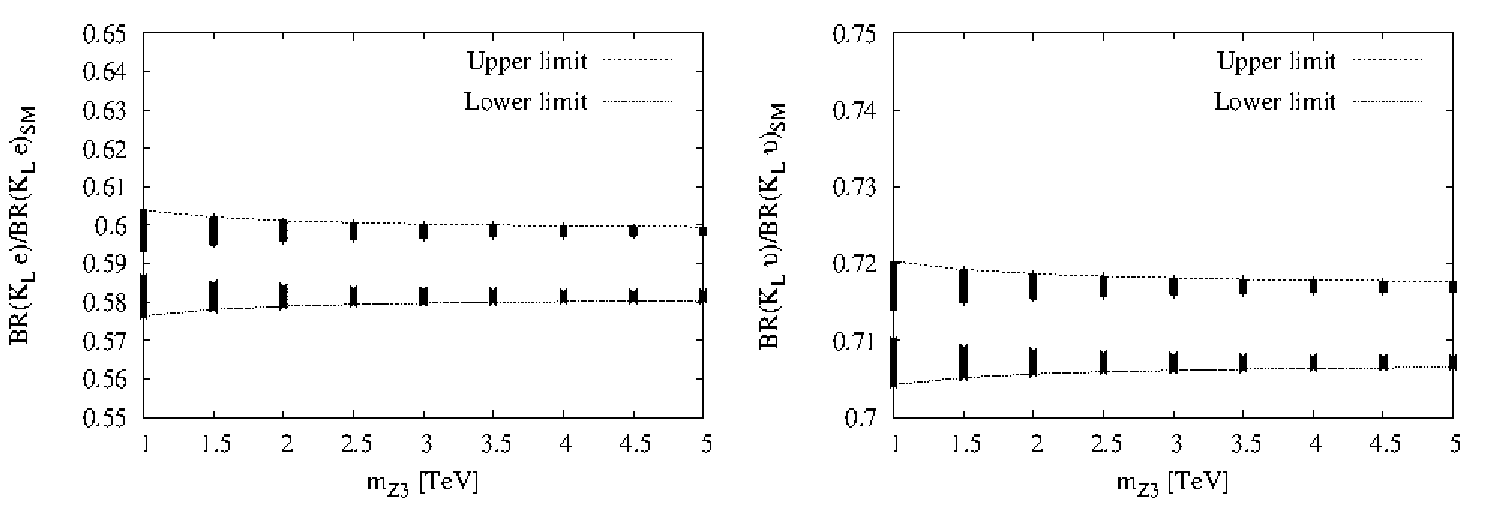} 
}
\caption{\label{fig:5b}Upper and lower bounds for the ratios ${\mbox{BR}(K_L \to \pi^0 e^+ e^-)}/{{\mbox{BR}(K_L \to \pi^0 e^+ e^-)}_{\rm SM}}$ (left) and ${\mbox{BR}(K_L \to \pi^0 \mu^+ \mu^-)}/{{\mbox{BR}(K_L \to \pi^0 \mu^+ \mu^-)}_{\rm SM}}$ (right), for different values of $m_{Z_3}$.}
\end{center}
\end{figure*}

\begin{figure} %Fig.7
\begin{center}
\resizebox{0.55\textwidth}{!}{
\includegraphics[scale=0.7]{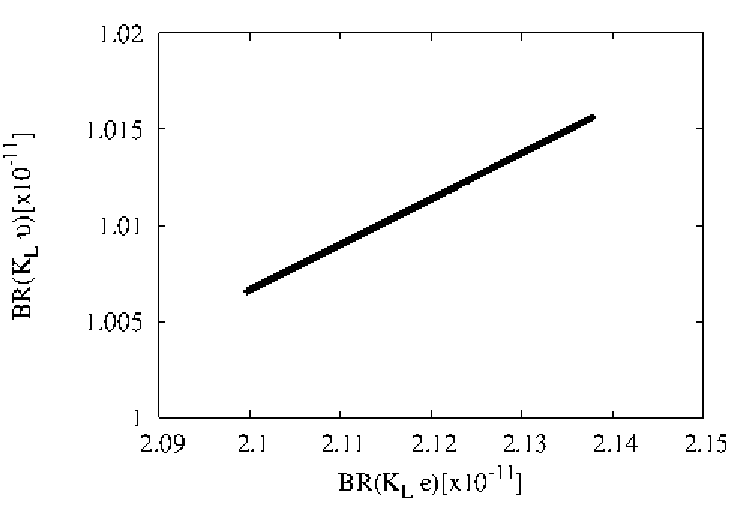} 
}
\caption{\label{fig:7}$\mbox{BR}(K_L \to \pi^0 \mu^+ \mu^-)$-$\mbox{BR}(K_L \to \pi^0 e^+ e^-)$ plane for $m_{Z_3}= 1$ TeV.}
\end{center}
\end{figure} 
 
%%%%%%%%%%%%%%%%%%%%%%%%%%%%%%%%%%%%%%%%%%%%%%%%%%%%%%%%%%%%%%%%%%%%%%%%%%%%%%%%%%%

\subsection{Implications for Rare Decays}
In order to study the implications of the obtained bounds for the modification in the rare decay amplitudes given in Section~\ref{sec:sec4}, we move over all the effective parameter space for each selected value of $m_{Z_3}$, and we calculate the amplitudes for the different decays. At difference with the analysis in Ref.~\cite{Promberger:2007py}, our procedure allows us to set not only upper bounds but also lower bounds on the corresponding BR as a function of the $Z_3$ mass. The present experimental data for these decays \cite{Nakamura:2010zzi} and the SM predictions \cite{Bu-Brod, Mescia:2006jd, Brod:2010hi} are given in Table~\ref{tab:4}. Our goal will be to determine these upper and lower bounds and establish their compatibility with the data in this Table.

For the decays $K^+ \to \pi^+ \bar \nu \nu$  and $K_{ L} \to \pi^0\nu\bar\nu$, using Eqs.~(\ref{kpvv}), (\ref{bklpn}), (\ref{X341}) and the effective parameter space, the bounds are given in Fig.~\ref{fig:3}. For the decay $K^+ \to \pi^+ \bar \nu \nu$ we see from Table~\ref{tab:4} that the experimental central value is greater than the SM prediction. Fig.~\ref{fig:3} shows that for low values of $m_{Z_3}$ the upper limit reaches the experimental central value, but for larger $m_{Z_3}$ both the upper and lower limits go closer to the SM prediction. For the decay $K_{L} \to \pi^0\nu\bar\nu$ we can see that big enhancements are expected for low $m_{Z_3}$. The bigger enhancements are expected for $m_{Z_3}$ values less that $\sim 2.5$ TeV. For the decay $K^+ \to \pi^+ \bar \nu \nu$ the greatest contributions come from the CP-conserving case and for $K_{ L} \to \pi^0\nu\bar\nu$ come from the CP-violating case in accordance with the setting in \cite{Promberger:2007py}, but with the difference that we have used all the effective parameter space which includes the CP-conserving, the CP-violating and the mixed cases.

To appreciate more clearly the departures from the SM, the ratios ${\mbox{BR}(K^+\rightarrow\pi^+\nu\bar\nu)}/{{\mbox{BR}(K^+\rightarrow\pi^+\nu\bar\nu)}_{\rm SM}}$ and ${\mbox{BR}(K_{ L}\rightarrow\pi^0\nu\bar\nu)}/{{\mbox{BR}(K_{ L}\rightarrow\pi^0\nu\bar\nu)}_{\rm SM}}$ vs. $m_{Z_3}$ are shown in Fig.~\ref{fig:3b}. For the decay $K^+ \to \pi^+ \bar \nu \nu$ we find enhancements around 2.8 times the SM prediction for low values of $m_{Z_3}$, but small departures from the SM for large $Z_3$ mass. For $K_{ L} \to \pi^0\nu\bar\nu$, enhancements of around 7.6 times the SM value are found for low values of $m_{Z_3}$. 

Comparing the upper and lower bounds on both decays we see that the possibility of enhancement is always larger than the possibility of suppression, except for large $m_{Z_3}$ where both of them are comparable.  

The plane $\mbox{BR}(K_{L} \to \pi^0\nu\bar\nu)-\mbox{BR}(K^+ \to \pi^+ \bar \nu \nu)$ is shown in Fig.~\ref{fig:3c}. The upper and lower branches correspond mainly to the CP-violating case and the horizontal branch to the CP-conserving case. It is implicit in this figure that the the mixed case is very restricted making the expressions in Eqs.~(\ref{X341}), (\ref{Y341}), (\ref{y_V}) and (\ref{y_A}) mainly real or imaginary. Notice that the pattern seen is similar to the one obtained in the Littlest Higgs model with T-parity in Ref.~\cite{Blanke:2006eb} and in the minimal 3-3-1 model in Ref.~\cite{Promberger:2007py}.

For the decays $B_s\to \mu^+\mu^-$ and $B_d\to \mu^+\mu^-$, using Eqs.~(\ref{blls}) and (\ref{Y341}), the limits are shown in Fig.~\ref{fig:4}. From the data in Table~\ref{tab:4} we see that, for all the selected range of values of $m_{Z_3}$, the upper bounds are smaller than the experimental upper limit. For the decay $B_d\to \mu^+\mu^-$ the upper and lower bounds are lower than the SM predictions.

${\mbox{BR}(B_s\to \mu^+\mu^-)}/{{\mbox{BR}(B_s\to \mu^+\mu^-)}_{\rm SM}}$ and ${\mbox{BR}(B_d\to \mu^+\mu^-)}/{{\mbox{BR}(B_d\to \mu^+\mu^-)}_{\rm SM}}$ vs. $m_{Z_3}$ are shown in Fig.~\ref{fig:4b}. For $B_s\to \mu^+\mu^-$ we see that both enhancements and suppressions are possible but both of them decrease with increasing values of $m_{Z_3}$ and are very small with respect to the SM prediction for $m_{Z_3}=5$ TeV. For $B_d\to \mu^+\mu^-$ suppression of around 0.87 times the SM value is possible for $m_{Z_3}\approx 1$ TeV, but for higher $m_{Z_3}$ the suppression can each 0.92 times the SM prediction.

Going back to Fig.~\ref{fig:4} it is interesting to note that, due to the small difference between the upper and lower bounds for ${\mbox{BR}(B_s\to \mu^+\mu^-)}$ and ${\mbox{BR}(B_d\to \mu^+\mu^-)}$, they are suppressed in this 3-4-1 extension even with the inclusion of new CP-violating phases, as compared with the case for the decays in Fig.~\ref{fig:3}. A similar behavior is reported in \cite{Promberger:2007py} for the minimal 3-3-1 model. It is important to remark that, as in the 3-3-1 model, a strong enhancement of this BR would rule out the extension we are studying.

We conclude analyzing the results for the decays $K_L \to \pi^0 e^+ e^-$ and $K_L \to \pi^0 \mu^+ \mu^-$. Using Eqs.~(\ref{y_V}), (\ref{y_A}) and (\ref{BrKpiLL}), the resulting limits are shown in Fig.~\ref{fig:5}. At this point it is important to clarify that the upper bound is built using the $+$ sign in Eqs.~(\ref{BrKpiLL}) and the lower bound is obtained using the $-$ sign. From the data in Table \ref{tab:4} we can see that these limits are lower than the SM predictions, but consistent with the upper experimental bounds. Notice also that the upper and lower limits are both almost independent of the $Z_3$ mass.

${\mbox{BR}(K_L \to \pi^0 e^+ e^-)}/{{\mbox{BR}(K_L \to \pi^0 e^+ e^-)}_{\rm SM}}$ and ${\mbox{BR}(K_L \to \pi^0 \mu^+ \mu^-)}/{{\mbox{BR}(K_L \to \pi^0 \mu^+ \mu^-)}_{\rm SM}}$ vs. $m_{Z_3}$ are shown in Fig.~\ref{fig:5b} from which we can see that these decays are suppressed compared to the SM prediction. Moreover, a comparison with Fig.~\ref{fig:4b} shows that, in the 3-4-1 extension, these BR are more suppressed than the ones for the $B_s\to \mu^+\mu^-$ and $B_d\to \mu^+\mu^-$ decays.

In the spirit of \cite{Isidori:2004rb}, we plot the plane ${\mbox{BR}(K_L \to \pi^0 \mu^+ \mu^-)}$-${\mbox{BR}(K_L \to \pi^0 e^+ e^-)}$ for $m_{Z_3}=1$ TeV in Fig.~\ref{fig:7}. This shows the correlation between these decays. It can be seen that they are very restricted. Comparing Fig.~\ref{fig:7} with the corresponding figures in Refs.~\cite{Isidori:2004rb, Mescia:2006jd}, it is noted that they are very close to the plots for the SM in these works. This fact can also be observed from a comparison between the two plots in Fig.~\ref{fig:5}.

\begin{figure*} %Fig.8
\begin{center}
\resizebox{0.98\textwidth}{!}{
\includegraphics[scale=1]{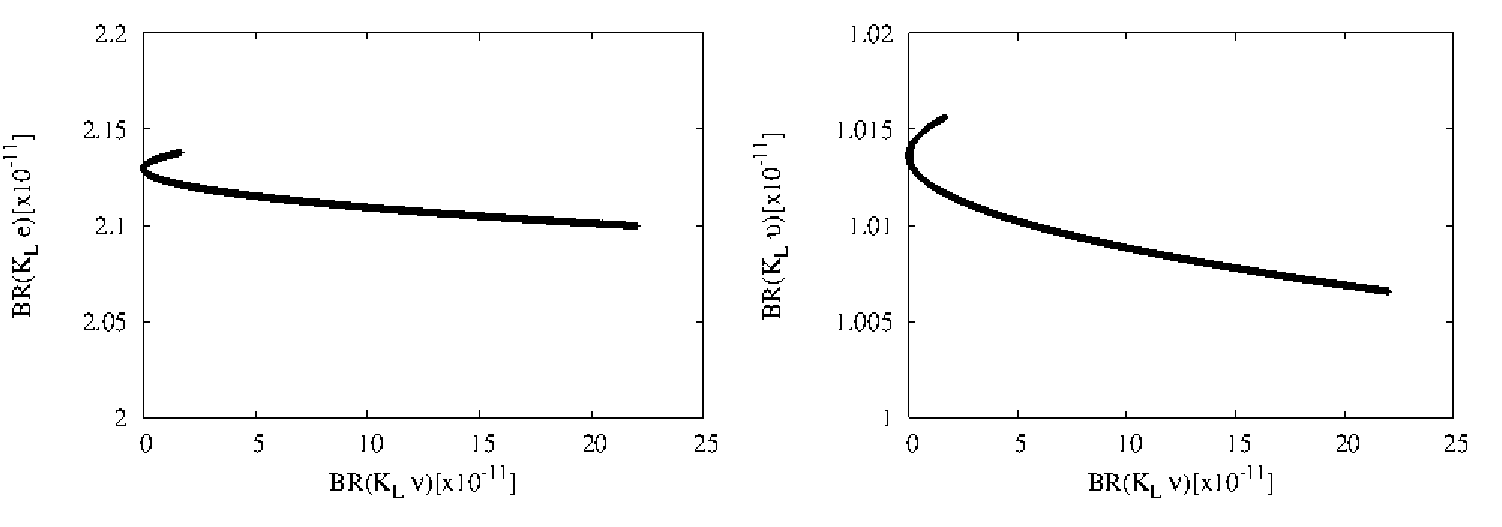} 
}
\caption{\label{fig:8} $\mbox{BR}(K_L \to \pi^0 e^+ e^-)$-$\mbox{BR}(K_{L} \to \pi^0\nu\bar\nu)$ plane (left) and $\mbox{BR}(K_L \to \pi^0 \mu^+ \mu^-)$-$\mbox{BR}(K_{L} \to \pi^0\nu\bar\nu)$ plane (right) for $m_{Z_3}= 1$ TeV.}
\end{center}
\end{figure*}

In Fig.~\ref{fig:8}, the planes $\mbox{BR}(K_L \to \pi^0 e^+ e^-)$-$\mbox{BR}(K_{L} \to \pi^0\nu\bar\nu)$ and $\mbox{BR}(K_L \to \pi^0 \mu^+ \mu^-)$-$\mbox{BR}(K_{L} \to \pi^0\nu\bar\nu)$ are shown. We find that, as it happens in many other models, the decays $K_L \to \pi^0 \mu^+ \mu^-$ and $K_L \to \pi^0 e^+ e^-$ are subject to weaker modifications than the decay $K_{L} \to \pi^0\nu\bar\nu$. We remark that this is contrary to the result found in \cite{Promberger:2007py} for the minimal 3-3-1 model. This result, then, can allow to test the 3-4-1 extension studied here when more accurate measurements of both decays become available.

\section{\label{sec:sec6}Summary}
In this work, in the framework of a 3-4-1 extension of the SM characterized by the values $b=1,\; c=-2$ in the most general expression for the electric charge generator and with chiral anomalies cancelling between the families, we have analyzed the effects of FCNC transmitted by one of the two extra neutral gauge bosons predicted by this construction in the case in which this new gauge boson $Z_3$ does not mix with the other two neutral gauge bosons. In this context we have studied in detail the constraints that well measured $\Delta F=2$ observables ($F=S,B$) impose on the parameter space associated to the $V_L^d$ matrix describing mixing in the down quark sector. This parameter space consists of five variables, namely the angles $\theta_{13}^d$, $\theta_{23}^d$ and the CP-violating phases $\phi'$, $\phi''$, related to the $V_L^d$ mixing matrix, and the $Z_3$ mass, and we have used the well measured quantities $\Delta M_K$, $\Delta M_{d/s}$, $\varepsilon_K$ and $\sin \Phi_d$.  For fixed values of some of the relevant $V_L^d$ parameters, allowed regions for the remaining parameters were plotted in Figs.~\ref{fig:1} and \ref{fig:2a} when $m_{Z_3}$ is allowed to vary in the typical range $1\,-\,5$ TeV. A calculation of the upper limits for $\theta_{13}$ and $\theta_{23}$, for different values of $m_{Z_3}$, is presented in Table \ref{tab:3} which shows that these bounds increase when $m_{Z_3}$ increases.  This allowed us to appreciate the behavior of the parameter space for selected values of the $Z_3$ mass.

We have then taken these results to obtain upper and lower bounds for the BR of several clean rare decay processes, namely $K^+ \to \pi^+ \bar \nu \nu$, $K_{L} \to \pi^0\nu\bar\nu$, $K_L \to \pi^0 l^+ l^-$ ($l=e, \mu$) and $B_{d/s}\to \mu^+ \mu^-$. We find sizeable enhancements and/or suppressions with respect to the SM prediction mainly for $m_{Z_3}$ values in the range $1\,-\,2$ TeV. In this range large enhancements, as compared with the suppressions, are predicted for the BR of the decays $K^+ \to \pi^+ \bar \nu \nu$ and $K_{ L} \to \pi^0\nu\bar\nu$. For $BR(B_{s}\to \mu^+ \mu^-)$ we find enhancements slightly larger than suppressions, while for $BR(B_{d}\to \mu^+ \mu^-)$ we find suppression for all the considered range of values of $m_{Z_3}$. Finally, for $BR(K_L \to \pi^0 l^+ l^-)$ we find upper and lower bounds smaller than the SM prediction but consistent with the upper experimental limit. These departures from the SM, presented in Figs.~\ref{fig:3b}, \ref{fig:4b} and \ref{fig:5b}, can be considered as signals of the 3-4-1 extension under consideration when looking at the data. From these plots is clear that the decays $K^+\rightarrow\pi^+\nu\bar\nu$ and $K_{ L}\rightarrow\pi^0\nu\bar\nu$ show the greatest differences with the SM predictions, but the decays $K_L \to \pi^0 l^+ l^-$ ($l=e,\mu)$ and $B_{d/s}\to \mu^+ \mu^-$ are very restricted in the 3-4-1 extension. It is important to note that, as in the minimal 3-3-1 model, a strong enhancement of the BR for the decays $B_{d/s}\to \mu^+ \mu^-$ would rule out the 3-4-1 extension studied here. 

The plane $\mbox{BR}(K_{L} \to \pi^0\nu\bar\nu-)-\mbox{BR}(K^+ \to \pi^+ \bar \nu \nu)$ in Fig.~\ref{fig:3c} allowed us to compare our results with the ones in the Littlest Higgs model with T-parity \cite{Blanke:2006eb} and with the ones in the minimal 3-3-1 model \cite{Promberger:2007py}, and it also helped us to better understand the structure of the expressions in Eqs.~(\ref{X341}), (\ref{Y341}), (\ref{y_V}) and (\ref{y_A}). It is important to remark that, contrary to what it is done in \cite{Promberger:2007py}, our results are obtained without the need of forcing a zero value neither for the real part nor for the imaginary part of these expressions. 

In Figs.~\ref{fig:7} and \ref{fig:8} we also plotted the planes ${\mbox{BR}(K_L \to \pi^0 \mu^+ \mu^-)}$-${\mbox{BR}(K_L \to \pi^0 e^+ e^-)}$, $\mbox{BR}(K_L \to \pi^0 e^+ e^-)$-$\mbox{BR}(K_{L} \to \pi^0\nu\bar\nu)$ and $\mbox{BR}(K_L \to \pi^0 \mu^+ \mu^-)$-$\mbox{BR}(K_{L} \to \pi^0\nu\bar\nu)$. It can be seen that the decays $K_L \to \pi^0 \mu^+ \mu^-$ and $K_L \to \pi^0 e^+ e^-$ are subject to weaker modifications than the decay $K_{L} \to \pi^0\nu\bar\nu$. When these plots are compared with similar ones obtained in other models, we find clear differences. In this sense, these results can allow to test the 3-4-1 extension under study here when more accurate measurements of the corresponding decays become available.

\section*{ACKNOWLEDGMENTS}
We acknowledge financial support from DIME at Universidad Nacional de Colombia-Sede Medell\'\i n.

\end{document}